\documentclass[12pt]{article}

\usepackage{amsfonts}
\usepackage{amsmath}
\usepackage{epsfig}
\usepackage{latexsym}
\usepackage{fancybox}
\usepackage{graphicx}
\usepackage{subfigure}
\usepackage{amssymb}
\numberwithin{equation}{section}
\allowdisplaybreaks

\def\spa#1{\phantom{\fbox{\rule[-#1cm]{0cm}{0cm}}}}

\def\be{\begin{equation}}
\def\ee{\end{equation}}
\def\bea{\begin{eqnarray}}
\def\eea{\end{eqnarray}}

\def\half{{1\over 2}}

\def\del{\partial}
\def\nn{\nonumber}

\renewcommand{\thefootnote}{\fnsymbol{footnote}}
\renewcommand{\thefootnote}{\fnsymbol{footnote}}

\def\[#1\]{\begin{align}#1\end{align}}

\begin{document}

\hfuzz=100pt
\title{{\Large \bf{
Giant graviton interactions \\
and\\ 
M2-branes ending on multiple M5-branes
}}}
\author{
Shinji Hirano$^{a,b}$\footnote{shinji.hirano@wits.ac.za} 
and
Yuki Sato$^c$\footnote{ysato.phys@gmail.com}
  \spa{0.5} 
\\
$^a${\small{\it School of Physics and Mandelstam Institute for Theoretical Physics}}
\\{\small{\it \& DST-NRF Centre of Excellence in Mathematical and Statistical Sciences (CoE-MaSS) }}
\\{\small{\it University of the Witwatersrand, WITS 2050, Johannesburg, South Africa}}\\ 
$^b${\small{\it Center for Gravitational Physics,
Yukawa Institute for Theoretical Physics}}
\\{\small{\it Kyoto University, Kyoto 606-8502, Japan }}\\
\\
$^c${\small{\it Department of Physics, Faculty of Science, Chulalongkorn University}}
\\ {\small{\it Thanon Phayathai, Pathumwan, Bangkok 10330, Thailand}}
\spa{0.3} 
}
\date{}

\maketitle
\vspace{-.8cm}
\centerline{}

\begin{abstract} 
We study splitting and joining interactions of giant gravitons with angular momenta $N^{1/2}\ll J\ll N$ in the type IIB string theory on $AdS_5 \times S^5$ by describing them as instantons in the tiny graviton matrix model introduced by Sheikh-Jabbari. At large $J$ the instanton equation can be mapped to the four-dimensional Laplace equation and the Coulomb potential for $m$ point charges in an $n$-sheeted Riemann space corresponds to the $m$-to-$n$ interaction process of giant gravitons.
These instantons provide the holographic dual of correlators of all semi-heavy operators and the instanton amplitudes exactly agree with the pp-wave limit of Schur polynomial correlators in ${\cal N}=4$ SYM computed by Corley, Jevicki and Ramgoolam. 

By making a slight change of variables
the same instanton equation is mathematically transformed into the Basu-Harvey equation 
which describes the system of M$2$-branes ending on M$5$-branes. As it turns out, the solutions to the sourceless Laplace equation on an $n$-sheeted Riemann space correspond to $n$ M5-branes connected by M2-branes and we find general solutions representing M2-branes ending on multiple M5-branes.
Among other solutions, the $n=3$ case describes an M2-branes junction ending on three M5-branes. The effective theory on the moduli space of our solutions might shed light on the low energy effective theory of multiple M5-branes.

\end{abstract}

\renewcommand{\thefootnote}{\arabic{footnote}}
\setcounter{footnote}{0}

\newpage

\section{Introduction}
\label{sec:introduction}

Giant gravitons are spherical branes moving fast along the great circle of the sphere in the $AdS_p\times S^q$ geometry \cite{McGreevy:2000cw, Grisaru:2000zn, Hashimoto:2000zp}
and correspond to Schur polynomial operators in dual CFTs \cite{Balasubramanian:2001nh, Corley:2001zk}. 
They form an orthogonal basis for multi-graviton states with Kaluza-Klein (KK) momenta and are appropriate objects for studying KK graviton interactions. In this paper we focus on giant gravitons in the type IIB string theory on $AdS_5\times S^5$ which is dual to ${\cal N}=4$ $U(N)$ SYM \cite{Maldacena:1997re}. 
On the CFT side, their interactions correspond to multi-point correlators of Schur polynomial operators and have been computed exactly for half-BPS giants in \cite{Corley:2001zk}. However, on the gravity side, being extended objects (spherical D3-branes), it is rather challenging to go beyond kinematics and study their dynamical interaction process except for so-called heavy-heavy-light three point interactions. 
This is the problem we tackle in the most part of this paper and we report modest but nontrivial progress on this issue. 

Instead of attempting to solve the issue once and for all, we consider a certain subset of giant gravitons, namely, those whose angular momentum $J$ are relatively small, 
\textit{i.e.} in the range $N^{1/2}\ll J\ll N$. These giants can be studied in the plane-wave background \cite{Berenstein:2002jq,Gubser:2002tv,Frolov:2002av}: 
For an observer moving fast in the sphere, the spacetime looks approximately like a plane-wave geometry.\footnote{
The plane-wave geometry can be obtained from $AdS_p\times S^q$ 
by taking the Penrose limit \cite{Penrose1976,Gueven:2000ru,Blau:2002dy}. 
} Thus if the size of giants is small enough,\footnote{Small giants are an oxymoron. They are small in the sense that their size is much smaller than the AdS radius, but they are not point-like and much larger than the Planck length.} the observer moving along with the giants can study them in the plane-wave background
\cite{Berenstein:2002jq,Gubser:2002tv,Frolov:2002av}.  

This strategy was inspired by the recent work of one of the authors which studied splitting and joining interactions of membrane giants in the M-theory on $AdS_4 \times S^7/ \mathbb{Z}_k$ at finite $k$ by zooming into the plane-wave background \cite{Kovacs:2015xha, Kovacs:2013una}. 
Since the M-theory on the plane-wave background is described by the BMN plane-wave matrix model \cite{Berenstein:2002jq},  
small membrane giants can be studied by this matrix quantum mechanics. Their idea is that since the vacua of the BMN matrix model represent spherical membranes, instantons interpolating among them correspond to the process of membrane interactions.
They explicitly constructed these instantons by mapping the BPS instanton equation \cite{Yee:2003ge} to Nahm's equation \cite{RBNahm} in the limit of large angular momenta where Nahm's equation becomes equivalent to the 3d Laplace equation \cite{Ward:1989nz,Hoppe:1994np}. The crux of their construction is to consider the Laplace equation not in the ordinary 3d Euclidean space but in a 3d analog of 2d Riemann surfaces, dubbed Riemann space \cite{Sommerfeld:1896, Sommerfeld:1899}.  
  
In our case of the type IIB string theory on $AdS_5\times S^5$, as it turns out, the most effective description of giant gravitons with the angular momentum $N^{1/2}\ll J\ll N$ is provided by the tiny graviton matrix model proposed by Sheikh-Jabbari \cite{SheikhJabbari:2004ik, SheikhJabbari:2005mf} rather than BMN's type IIB string theory on the pp-wave background.\footnote{In this paper we refer to the tiny graviton matrix model as the type IIB plane-wave matrix model.} 
The description of giant graviton interactions is similar to the above M-theory case, and in the large $J$ limit the instanton equation in this matrix quantum mechanics can be mapped to the Laplace equation but in four dimensions instead of three.
As we will see, the 4d Coulomb potential for $m$ point charges in an $n$-sheeted Riemann space corresponds to the $m$-to-$n$ interaction process of giant gravitons.
An advantage over the M-theory case is that we can compare our description of giant graviton interactions to that of ${\cal N}=4$ SYM. Indeed, we find that the instanton amplitude exactly agrees with the pp-wave limit of Schur polynomial correlators in ${\cal N}=4$ SYM computed by Corley, Jevicki and Ramgoolam \cite{Corley:2001zk}. This also implies that these instantons successfully provide the holographic dual of correlators of all semi-heavy operators.

Last but not the least, as a byproduct of this study we are led to find new results on elusive M5-branes. 
By a slight change of variables, the instanton equation of the type IIB plane-wave matrix model is identical to the Basu-Harvey equation which describes the system of M$2$-branes ending on M$5$-branes \cite{Basu:2004ed}.
In the large $J$ limit which corresponds, in the Basu-Harvey context, to a large number of M2-branes, we find the solutions describing M2-branes ending on multiple M5-branes, including the funnel solution \cite{Howe:1997ue, Niarchos:2012pn} and an M2-branes junction connecting three M5-branes as simplest examples. The number $n$ of M5-branes corresponds to the number of sheets in the Riemann space, and  
somewhat surprisingly, multiple M5-branes solutions are constructed from a trivial constant electrostatic potential.
Upon further generalisations, the effective theory on the moduli space of our solutions might shed light on the low energy effective theory of multiple M5-branes \cite{Ho:2008nn, Lambert:2010wm, Chu:2009iv, Saemann:2017zpd}.

This paper is organised as follows:  In Section \ref{sec:iibplanewavematrixmodel}, we review the IIB plane-wave matrix model and 
its BPS vacua  which contain concentric fuzzy three-spheres. 
We then discuss the instanton equation and find the (anti-)instanton action for the $m$-to-$n$ joining and splitting process of giant gravitons. As the first check of our proposal we show that the instanton amplitude $e^{-S_E}$ in the case of the $2$-to-$1$ interaction agrees with the 3-point correlators of antisymmetric Schur operators in the dual CFT, 
\textit{i.e.} ${\cal N}=4$ SYM. 
In Section \ref{sec:4dlaplaceequationinriemannspaces}, we transform the instanton equation to  the Basu-Harvey equation by a suitable change of variables and show that in the large $J$ limit it is further mapped locally to the 4d Laplace equation. 
We then solve the 4d Laplace equation in multi-sheeted Riemann spaces and find the solutions which describe the generic $m$-to-$n$ joining and splitting process of (concentric) sphere giants. 
In Section \ref{sec:giantgravitoncorrelatorsincft}, we discuss the pp-wave limit of correlators of antisymmetric Schur operators in the dual CFT and show that they exactly agree with the instanton amplitudes obtained in Section \ref{sec:4dlaplaceequationinriemannspaces}.
In Section \ref{sec:m2m5basuharveyequation}, we study the Basu-Harvey equation in the original context, namely, as a description of the M2-M5 brane system. In the large $J$ limit corresponding to a large number of M2-branes, we find the solutions to the 4d Laplace equation which describe M2-branes ending on multiple M5-branes. 
Section \ref{sec:discussion} is devoted to summary and discussions. 
In the appendices \ref{sec:ap:derivationoflaplaceequation}, \ref{sec:ap:euclideanthreebranetheory}  and  
\ref{sec:ap:threespheres} we elaborate further on some technical details.


\section{IIB plane-wave matrix model}
\label{sec:iibplanewavematrixmodel}

The tiny graviton matrix model was proposed by Sheikh-Jabbari 
as a candidate for the discrete lightcone quantisation (DLCQ) of the type IIB string theory 
on the maximally supersymmetric ten-dimensional plane-wave background \cite{SheikhJabbari:2004ik}. 
We refer to this matrix model as the IIB plane-wave matrix model in this paper.


Here we outline the derivation of the IIB plane-wave matrix model. 
The bosonic part of the IIB plane-wave matrix model can be obtained
by a matrix regularisation of the effective action for a 3-brane \cite{SheikhJabbari:2004ik}:
\[
S 
= -T \int \text{d}t \text{d}^3\sigma\ 
\left( 
\sqrt{ |\det (h_{\mu \nu})| } 
+ C_{\hat \mu \hat \nu \hat \rho \hat \lambda}\   
\frac{\partial x^{\hat \mu}}{\partial t} 
\frac{\partial x^{\hat \nu}}{\partial \sigma^1}
\frac{\partial x^{\hat \rho}}{\partial \sigma^2}
\frac{\partial x^{\hat \lambda}}{\partial \sigma^3}
\right)\ , 
\label{eq:dbipluscsaction}
\]
where 
$T=1/((2\pi)^3 g_s l^4_s)$ is the D3-brane tension with 
$g_s$ and $l_s$ being the string coupling constant and string length, respectively. 
The world-volume coordinates are $\sigma^{\mu} = (t, \sigma^l)$ 
with $\mu = 0,1,2,3$ and $l=1,2,3$. 
The indices for the target space are hatted, 
$\hat \mu,\hat \nu,\hat \rho,\hat \lambda = +,-,1,\cdots,8$. 
The background metric is the plane-wave geometry: 
\[
g_{\hat \mu \hat \nu} \text{d}x^{\hat \mu} \text{d}x^{\hat \nu} 
= - 2 \text{d}x^+ \text{d}x^- 
- \mu^2 (x^ix^i + x^ax^a) \text{d}x^+ \text{d}x^+ 
+\text{d}x^i\text{d}x^i 
+ \text{d}x^a\text{d}x^a\ ,
\label{eq:planewave}
\]  
with $i=1,2,3,4$ and $a=5,6,7,8$. 
The induced metric on the 3-brane is 
\[
h_{\mu \nu} 
= g_{\hat \mu \hat \nu} \partial_{\mu} x^{\hat \mu} \partial_{\nu} x^{\hat \nu}\ ,
\label{eq:inducedmetric} 
\] 
and $C_{\hat \mu \hat \nu \hat \rho \hat \sigma}$ is the Ramond-Ramond 4-form with nonvanishing components
\[
C_{+ijk} = - \mu \epsilon_{ijkl} x^l\ , 
\ \ \ C_{+abc} = - \mu \epsilon_{abcd} x^d\ . 
\label{eq:rr4}
\] 
The parameter $\mu$ in (\ref{eq:planewave}) and (\ref{eq:rr4}) is the mass parameter. 

In the lightcone gauge we fix $x^+ =t$ while imposing  
$h_{0l}=0$ and choose the spatial world-volume coordinates $\sigma^l$ 
such that the lightcone momentum density $-p_-$ is a constant. 
The lightcone Hamiltonian for the 3-brane 
is then given by \cite{SheikhJabbari:2004ik, Sadri:2003mx}
\[
-P_+
&= \int \text{d}^3\sigma\ 
\biggl[ 
\frac{[\sigma]}{2(-P_-)} (p^I)^2 
+ \frac{\mu^2 (-P_-)}{2[\sigma]} (x^I)^2 
+ \frac{T^2 [\sigma]}{2 \cdot 3! (-P_-) } 
\{x^I,x^J,x^K \}^2 \notag \\
& \ \ \ 
- \frac{\mu T}{3!} 
\left( 
\epsilon^{ijkl} x^i \{ x^j,x^k,x^l \} 
+ \epsilon^{abcd} x^a \{ x^b,x^c,x^d \} 
\right)
\biggl]\ , 
\label{eq:lightconehamiltonian}
\] 
where
$I,J,K=1,2,\cdots,8$ are transverse directions, 
$x^I=(x^i,x^a)$ and $p^I=(p^i,p^a)$ are the conjugate momenta of $x^I$. 
$P_{\pm}$ are the zero-modes of $p_{\pm}$ and the conjugate momenta of $x^{\pm}$.   
$[\sigma]$ is the total volume in the $\sigma$-space defined as 
\[
[\sigma] = \int \text{d}^3 \sigma\ .
\label{eq:sigmavolume}
\]  
The Nambu three-bracket in (\ref{eq:lightconehamiltonian}) 
is defined for real functions, 
$f_p (\sigma)$ with $p=1,2,3$,   
as
\[
\{ f_1, f_2,f_3 \} 
= \epsilon^{lmn} 
\frac{\partial f_1}{\partial \sigma^{l}} 
\frac{\partial f_2}{\partial \sigma^{m}} 
\frac{\partial f_3}{\partial \sigma^{n}}\ . 
\label{eq:nambuthreebracket}
\]  
Since the constraints, $h_{r0}=0$, 
can be recast as 
\[
\frac{\partial x^-}{\partial \sigma^r} = \frac{[\sigma]}{(-p_-)} p_I \frac{\partial x^I}{\partial \sigma^r}\ ,
\label{eq:h0ris0}
\]
the dynamics of $x^-$ can be determined by that of the transverse directions. 
The constraints (\ref{eq:h0ris0}) together with 
the conditions,  
$\epsilon^{lmn} \frac{\partial}{\partial \sigma^m}\frac{\partial}{\partial \sigma^n}x^-
=0$, 
can be rewritten as 
\[
\epsilon^{lmn} 
\frac{\partial x^I}{\partial \sigma^m}
\frac{\partial p_I}{\partial \sigma^n}
=0\ . 
\label{eq:volumepreservingdiff}
\]
This should correspond to the generator of the residual local symmetry  
analogous to the area-preserving diffeomorphism of 
the membrane theory in the lightcone gauge.  

We further compactify the $x^-$ in the background (\ref{eq:planewave}) 
on a circle of radius $R$, 
resulting in the quantised total lightcone momentum:
\[
-P_-  = -p_- [\sigma] = \frac{J}{R}\ , 
\label{eq:totallightconemomentum} 
\] 
where $J$ is an integer. 

We replace the functions by matrices, 
\[
x^I(\sigma) \ \ \  &\to \ \ \   X^I\ ,
\label{eq:matrixregularisation1} 
\\ 
p^I(\sigma) \ \ \  &\to \ \ \   \frac{J}{[\sigma]}\ \Pi^I\ , 
\label{eq:matrixregularisation2}
\]
where $X^I$ and $\Pi^I$ are $J\times J$ matrices, 
and implement the further replacements, 
\[
\{x^I,x^J,x^K\} \ \ \  &\to \ \ \   \frac{1}{(il)^2} [X^I,X^J,X^K,\Upsilon_5]\ , 
\label{eq:matrixregularisation3}
\\
\frac{1}{[\sigma]}\int \text{d}^3\sigma\ \ast \ \ \  &\to \ \ \  \frac{1}{J} \text{tr}\ \ast\ ,  
\label{eq:matrixregularisation4}
\]
where $\Upsilon_5$ is a non-dynamical $J \times J$ matrix explained in Appendix \ref{sec:ap:threespheres} 
and the quantum Nambu four-bracket is defined among matrices, 
$F_p$ with $p=1,2,3,4$, as 
 \[
[F_1,F_2,F_3,F_4]  
= \frac{1}{4!} \ \epsilon^{pqrs} F_p F_q F_r F_s\ . 
\label{eq:quantumnambufourbracket} 
 \]
In (\ref{eq:matrixregularisation3}) the parameter $l$ is analogous to 
$\hbar$ in quantum mechanics and given by\footnote{
We explain how to fix the parameter $l$ in Appendix \ref{sec:ap:threespheres}.
}  
\[
l= \sqrt{\frac{[\sigma]}{2\pi^2 J}}\ . 
\label{eq:hbariniib}
\] 
With these replacements (\ref{eq:matrixregularisation1}) - (\ref{eq:matrixregularisation4}) 
we finally obtain the bosonic part of the lightcone Hamiltonian of 
the IIB plane-wave matrix model \cite{SheikhJabbari:2004ik},\footnote{
The bosonic lightcone Hamiltonian (\ref{eq:bosonichamiltonian}) becomes 
the one in \cite{SheikhJabbari:2004ik} by choosing the unit, $4\pi l^4_s=1$, and
changing $\mu \to - \mu$. 
This sign difference originates from that in the replacement (\ref{eq:matrixregularisation3}). 
}
\[
H_B &= 
R\ \text{tr} 
\biggl[ 
\frac{1}{2} (\Pi^I)^2 
+ \frac{1}{2} \left( \frac{\mu}{R} \right)^2 (X^I)^2 
+ \frac{(2\pi^2 T)^2}{2 \cdot 3!}   
[X^I,X^J,X^K,\Upsilon_5]^2 \notag \\
& \ \ \ 
+ \frac{2\pi^2 \mu T}{3! R } 
\left(
\epsilon^{ijkl} X^i [X^j,X^k,X^l,\Upsilon_5] 
+ \epsilon^{abcd} X^a [X^b,X^c,X^d,\Upsilon_5]
\right)
\biggl]\ .
\label{eq:bosonichamiltonian}
\] 
The full supersymmetric IIB plane-wave matrix model with 
$PSU(2|2)\times PSU(2|2) \times U(1)$ symmetry is given 
by the following lightcone Hamiltonian \cite{SheikhJabbari:2004ik}:
\[
H &
= H_B 
 + R \ \text{tr} 
\biggl[ 
\frac{\mu}{R} 
\left(
\Psi^{\dagger \alpha \beta} \Psi_{\alpha \beta} 
- \Psi_{\dot \alpha \dot \beta} \Psi^{\dot \alpha \dot \beta}
\right) \notag \\
& \ \ \ 
-2  \left( 2\pi^2 T \right)
\left(
\Psi^{\dagger \alpha \beta} (\sigma^{ij})_{\alpha}{}^{\delta} [X^i,X^j,\Psi_{\delta \beta},\Upsilon_5] 
+ \Psi^{\dagger \alpha \beta} (\sigma^{ab})_{\alpha}{}^{\delta} [X^a,X^b,\Psi_{\delta \beta}, \Upsilon_5]
\right) \notag \\
& \ \ \ 
+ 2  \left( 2\pi^2 T \right) 
\left(
\Psi_{\dot \delta \dot \beta} (\sigma^{ij})_{\dot \alpha}{}^{\dot \delta} [X^i,X^j,\Psi^{\dagger \dot \alpha \dot \beta},\Upsilon_5] 
+ \Psi_{\dot \delta \dot \beta} (\sigma^{ab})_{\dot \alpha}{}^{\dot \delta} [X^a,X^b,\Psi^{\dagger \dot \alpha \dot \beta},\Upsilon_5]
\right)
\biggl]\ ,
\label{eq:fullhamiltonian}
\]
where $H_B$ is given by (\ref{eq:bosonichamiltonian}). 
The $J\times J$ matrices $\Psi$ 
are spinors of two $SU(2)$'s and 
each spinor carries two kinds of indices   
in which each index is the Weyl index of one of two $SO(4)$'s    
under the isomorphism,    
$SO(4) \cong SU(2) \times SU(2)$.  
There exist the constraints which would be a matrix regularisation 
of the supersymmetric extension 
of (\ref{eq:volumepreservingdiff}) in the continuum theory:   
\[
i[X^i,\Pi^i] 
+ i[X^a,\Pi^a] 
+ 2\Psi^{\dagger \alpha \beta} \Psi_{\alpha \beta} 
+2\Psi^{\dagger \dot \alpha \dot \beta} \Psi_{\dot \alpha \dot \beta} 
\approx 0\ , 
\label{eq:gausslaw}
\]    
on the physical states \cite{SheikhJabbari:2004ik}. 
The bracket $[\ , \ ]$ denotes the matrix commutator.   
The lightcone Hamiltonian (\ref{eq:fullhamiltonian}) can be derived from 
a Lagrangian of the corresponding supersymmetric matrix quantum mechanics 
with $U(J)$ gauge symmetry in which the component of the gauge field 
$A_0$ is set to zero. 
In order to maintain this gauge condition along the lightcone time flow, 
one has to impose the Gauss-law constraints which are nothing but (\ref{eq:gausslaw}). 
The $PSU(2|2)\times PSU(2|2)\times U(1)$ superalgebra in the plane-wave background 
can be realised by the $J\times J$ matrices \cite{SheikhJabbari:2004ik}.

The plane-wave background (\ref{eq:planewave}) can be obtained 
from $AdS_5 \times S^5$: One starts with the global $AdS_5 \times S^5$ spacetime  
\[
\text{d}s^2 
= R^2_{S} 
\left[ 
- \cosh^2 \rho \text{d}\tau^2 
+ \text{d} \rho^2 
+ \sinh^2 \rho \text{d} \Omega^2_3 
+ \text{d} \theta^2_5
+ \sum^{9}_{n=6} \left( \prod^{n-1}_{m=5} (\sin \theta_m)^2 \text{d}\theta^2_n \right) 
\right]\ ,
\label{eq:ads5timess5}
\]   
where $R_{S}$ denotes the $AdS_5$ and $S^5$ radius of curvature. 
One then zooms into the trajectory of a particle moving along a great circle in $S^5$ 
at large angular momentum $J$ and sitting at the centre $\rho=0$ of $AdS_5$. 
To see what happens, one introduces rescaled coordinates,
\[
\tau = \frac{x^0}{R_{S}}\ , \ \ \ \rho = \frac{\sqrt{(x^i)^2}}{R_{S}}\ , \ \ \ 
\theta_9 = \frac{x^9}{R_{S}}\ , \ \ \ \theta_a = \frac{\pi}{2} + \frac{x^a}{R_{S}}\ , 
\label{eq:coorinatechangeadstoppwave}
\] 
where $i=1,2,3,4$ and $a=5,6,7,8$ and further introduces the lightcone coordinates 
\[
x^+ = \frac{1}{2\mu_0} \left( x^0 + x^9 \right)\ , \ \ \ 
x^- = \mu_0 ( x^0 - x^9) \ , 
\label{xpm}
\]
with $\mu_0$ being a dimensionless parameter. 
Due to the strong centrifugal force, at large angular momentum $J=-i\partial_{\theta_9}$ 
the trajectory of a particle is confined to the region close to 
the great circle in the $56$ plane of $\mathbb{R}^6$ where $S^5$ is embedded. This implies that
\[
\frac{|x^a|}{R_{S}} \ll 1, \ \ \ \frac{|x^-|}{R_{S}} \ll 1\ . 
\label{eq:conditionppwave1} 
\]  
Since the particle at the centre of $\rho =0$ of $AdS_5$, 
we also have 
\[
\frac{|x^i|}{R_{S}} \ll 1\ . 
\label{eq:conditionppwave2}
\]
In this region of spacetime, (\ref{eq:conditionppwave1}) and (\ref{eq:conditionppwave2}), 
the global $AdS_5 \times S^5$ spacetime (\ref{eq:ads5timess5}) is approximated by 
the plane-wave background (\ref{eq:planewave}) with the identification
\[
\mu = \frac{\mu_0}{R_{S}}\ . 
\label{eq:mutor}
\] 
The relation between $R$ and $R_{S}$ is given by
\[
R = \mu_0 R_S\ ,
\label{eq:relationamongrandrs}
\]
because 
\[
-P_- 
= 
\frac{1}{2\mu_0} 
\left(
-P_0 + P_9
\right) 
\approx \frac{1}{\mu_0}P_9 
= \frac{J}{\mu_0 R_S}\ .
\label{eq:pminusandp9}
\]
In this paper, the plane-wave background is the approximation of  
the $AdS_5 \times S^5$ geometry near the observer with large angular momentum $J$.  
Thus the matrix size $J$ in the IIB plane-wave matrix model 
is considered to be very large for our purposes.

\subsection{Vacua}
\label{sec:zeroenergysolutions}
Similar to the plane-wave matrix model for M-theory \cite{Berenstein:2002jq}, 
the IIB plane-wave matrix model has abundant static zero energy configurations \cite{SheikhJabbari:2004ik}. 
Since the bosonic Hamiltonian (\ref{eq:bosonichamiltonian}) can be expressed as a sum of squares, 
\[
H_B 
&= \frac{R}{2} \text{tr} 
\biggl[
(\Pi^I)^2 
+ 
\frac{\left( 2\pi^2 T \right)^2}{2}  
\left( 
[X^i,X^a,X^b,\Upsilon_5]^2
+[X^a,X^i,X^j,\Upsilon_5]^2
\right) \notag \\
& \ \ \ +
\left(
\frac{\mu}{R}X^i + \frac{2\pi^2 T}{3!}  
\epsilon^{ijkl}[X^j,X^k,X^l,\Upsilon_5]
\right)^2 \notag \\
& \ \ \ +
\left(
\frac{\mu}{R}X^a + \frac{2\pi^2 T}{3!} 
\epsilon^{abcd}[X^b,X^c,X^d,\Upsilon_5]
\right)^2
\biggl]\ ,
\label{eq:bosonichamiltonian2}
\]    
there exist three kinds of vacua \cite{SheikhJabbari:2004ik}: 
\[
X^i &= - \frac{2\pi^2 R T}{3! \mu}  
\epsilon^{ijkl}[X^j,X^k,X^l,\Upsilon_5] \ne 0\ , \ \ \ 
X^a=0\ , \label{eq:adsgiant} \\
X^a &= - \frac{2\pi^2 R T}{3! \mu} 
\epsilon^{abcd}[X^b,X^c,X^d,\Upsilon_5] \ne 0\ , \ \ \ 
X^i=0\ , \label{eq:s5giant} \\
X^a &= X^i =0\ . \label{eq:trivial}
\]
The solutions to (\ref{eq:adsgiant}) and (\ref{eq:s5giant}) preserve a half of the supersymmetries and represent concentric fuzzy $S^3$ classified by $J\times J$ representations of $Spin (4) = SU(2)_L \times SU(2)_R$ \cite{SheikhJabbari:2004ik, SheikhJabbari:2005mf}. (See Appendix \ref{sec:ap:threespheres} for more details.)
These fuzzy $S^3$'s are identified with giant gravitons 
and in particular the solutions to (\ref{eq:adsgiant}) and (\ref{eq:s5giant}) 
are called $AdS$ and sphere giants, respectively. 
For irreducible representations of $Spin(4)$, the solutions to (\ref{eq:adsgiant}) and (\ref{eq:s5giant}) 
become a single giant graviton with the radius
\[
r = \sqrt{\frac{\mu J}{2\pi^2 RT}}=R_S \sqrt{\frac{J}{N}}
\ ,  
\label{eq:radiusofgiant}
\] 
which can be inferred from 
(\ref{eq:mutor}), 
(\ref{eq:relationamongrandrs}) and 
\[
R^4_S = 4\pi N g_s l^4_s \ . 
\label{eq:adscftdictionary}
\] 
We denote this $J\times J$ irreducible representation 
by $\mathbb{J}$. As for reducible representations, 
the matrices are block-diagonal and each size is, 
say, $J_l$ with $l=1,2,\cdots,n$ and $J=J_1+J_2 + \cdots + J_n$, 
which can be expressed as $\mathbb{J}_1 \oplus \mathbb{J}_2 \oplus \cdots \oplus \mathbb{J}_n$. 
This configuration corresponds to the concentric $n$ fuzzy $S^3$'s 
and the block of size $J_l$ has the radius,
\[
r_l = \sqrt{\frac{\mu J_l}{2\pi^2 RT}}
= R_S \sqrt{\frac{J_l}{N}}\ .  
\label{eq:blockradiusofgiant}
\] 
In order for the plane-wave approximation to be valid, 
the radius of each giant graviton $r_l$ 
should be much smaller than $R_S$. 
This leads to the condition 
\[
J_l \ll N \ . 
\label{eq:upperboundofj}
\]
Quantum corrections are well controlled 
if the length scale $r_l$ is much 
larger than the 10d Planck length 
$l_p=g^{1/4}_s l_s$. This yields another condition
\[
N^{1/2} \ll J_l\ .  
\label{eq:lowerboundofj}
\]   
Combining the two (\ref{eq:upperboundofj}) and (\ref{eq:lowerboundofj}), 
we obtain the bound for $J_l$:
\[
N^{1/2} \ll J_l \ll N \ .
\label{eq:boundofj}
\]

In the following, we study the tunnelling processes which interpolate various vacua (corresponding to giant gravitons) classified 
by the representation of $Spin(4)$, {\it i.e.} (anti-)instanton solutions of the IIB plane-wave matrix model. 
As will be elaborated further, the (anti-)instantons describe splitting or joining interactions 
of concentric giants.\footnote{These vacua are $1/2$-BPS and marginally stable. Nonetheless, the instanton and anti-instanton amplitudes corresponding, respectively, to splitting and joining interactions are nonvanishing. However, they are equal and there is an equilibrium of splitting and joining processes.} 
Similar (anti-)instantons have been discussed
in the BMN matrix model \cite{Yee:2003ge, Kovacs:2015xha} 
and our analysis will be analogous to theirs.

\subsection{Instanton equations}
\label{sec:instantonequations}

In order to find (anti-)instanton solutions, 
we consider the Euclidean IIB plane-wave matrix model.  
Hereafter we ignore the fermionic matrices by setting $\Psi=0$. 
The Euclidean action for the bosonic IIB plane-wave matrix model is  
\[
S_E 
&= \frac{1}{2R} \text{tr}\int \text{d}t\ 
\biggl[
\left( \frac{\text{d} X^I}{\text{d}t}  \right)^2 
+ \mu^2 (X^I)^2 
+ \frac{(2\pi^2 R T)^2}{3!} 
[X^I,X^J,X^K,\Upsilon_5]^2 \notag \\
& \ \ \ 
+ \frac{2\pi^2 \mu R T}{3}   
\left(
\epsilon^{ijkl} X^i [X^j,X^k,X^l,\Upsilon_5] 
+ \epsilon^{abcd} X^a [X^b,X^c,X^d,\Upsilon_5]
\right)
\biggl]\ , 
\label{eq:iibbosoniceuclideanaction}
\]
where $t$ is now the Euclidean time. 
One can show that the Euclidean action (\ref{eq:iibbosoniceuclideanaction}) 
can be rewritten as sum of squares and boundary terms:
\[
S_E 
= \frac{1}{2R} \text{tr} \int \text{d}t\ 
\biggl[
&\left( 
\frac{\text{d} X^i}{\text{d} t} \pm \mu X^i \pm  \frac{2\pi^2 RT}{3!} \epsilon^{ijkl} [X^j,X^k,X^l,\Upsilon_5]
\right)^2 \notag \\
+&\left( 
\frac{\text{d} X^a}{\text{d} t} \pm \mu X^a \pm  \frac{2\pi^2 RT}{3!} \epsilon^{abcd} [X^b,X^c,X^d,\Upsilon_5]
\right)^2 \notag \\
+&\frac{(2\pi^2 R T)^2}{2} 
\left( 
[X^i,X^a,X^b,\Upsilon_5]^2 
+ [X^a,X^i,X^j,\Upsilon_5]^2
\right) \notag \\
\mp & \frac{\text{d}}{\text{d}t} 
\left( 
\mu (X^i)^2 + \frac{2\pi^2 RT}{12} \epsilon^{ijkl} X^i [X^j,X^k,X^l,\Upsilon_5]
\right) \notag \\
\mp & \frac{\text{d}}{\text{d}t} 
\left( 
\mu (X^a)^2 + \frac{2\pi^2 RT}{12} \epsilon^{abcd} X^a [X^b,X^c,X^d,\Upsilon_5]
\right)
\biggl]\ .
\label{eq:iibbosoniceuclideanaction2}
\]
Therefore, the Euclidean action is bounded by the boundary terms 
and (anti-)instantons are configurations which saturate the bound. 
In this manner, 
the (anti-)instanton equations can be obtained:
\[
\frac{\text{d} X^i}{\text{d} t} 
\pm \mu X^i 
\pm \frac{2\pi^2 RT}{3!} \epsilon^{ijkl} [X^j,X^k,X^l,\Upsilon_5] = 0\ , 
\ \ \ X^a=0\ , 
\label{eq:instantoneqi}
\]
and the same equations with the replacement, $(i,j,k,l) \leftrightarrow (a,b,c,d)$.  
We will focus on the (anti-)instanton equation (\ref{eq:instantoneqi}) 
associated with $AdS_5$, but the $S^5$ case can be obtained 
from the $AdS_5$ case by interchanging the indices. 
One notices that the (anti-)instanton equation (\ref{eq:instantoneqi}) 
implies the equation:
\[
\frac{\text{d}}{\text{d}t} W[X] 
= \mp \frac{1}{2} \biggl| \frac{\partial W[X]}{\partial X^i} \biggl|^2\ ,
\label{eq:gradientflow}
\]
where the double sign is correlated with the one in (\ref{eq:instantoneqi}) 
and 
\[
W[X] = \mu (X^i)^2 
+ \frac{2\pi^2 RT}{12} \epsilon^{ijkl} 
X^i [X^j,X^k,X^l,\Upsilon_5]\ . 
\label{eq:wofx}   
\]
The equation (\ref{eq:gradientflow}) implies that 
the functional $W[X]$ monotonically decreases or increases 
in progress of the Euclidean time depending on a choice of the double sign. 
We call solutions such that $W[X]$ decreases (increases) instantons (anti-instantons). 
These tunnelling processes would be governed by the path integral with boundary conditions: 
\[
X^j (-\infty) = X^j_0 (-\infty)\ , \ \ \ 
X^j (\infty) = U X^j_0 (\infty) U^{-1}\ , 
\label{eq:pathintegralboundaryconditions}
\]   
where $X^j_0 (\pm \infty)$ are matrices forming static concentric fuzzy $S^3$'s 
and $U$ is an arbitrary unitary matrix introduced 
to maintain the gauge condition, $A_0=0$.

Using the equation (\ref{eq:gradientflow}), 
one can show that the (anti-)instanton action is non-negative: 
\[
S_E 
= \mp \frac{1}{2R} \text{tr} \ W[X(t)]\biggl|^{\infty}_{-\infty} 
= \mp \frac{\mu}{4R} \text{tr} \ ((X^i_0 (\infty))^2 - (X^i_0 (-\infty))^2) \ge 0\ .  
\label{eq:positivity}
\]   
In particular,  we are going to consider instantons interpolating between the vacuum of $m$ giant gravitons, 
$\mathbb{J}_1 \oplus \mathbb{J}_2 \oplus \cdots \oplus \mathbb{J}_m$, 
at $t=-\infty$ 
and that of $n$ giant gravitons, 
$\mathbb{J'}_1 \oplus \mathbb{J'}_2 \oplus \cdots \oplus \mathbb{J'}_n$, 
at $t=+\infty$, 
where $J=J_1+J_2+\cdots + J_m = J'_1+J'_2+\cdots + J'_n$.  
The Euclidean action in this case becomes
\[
S_E 
= - \frac{1}{4N} 
\left( 
\sum^n_{i=1} J'^2_i - \sum^m_{j=1} J^2_j
\right)\ .
\label{eq:instantonactionntom}
\]
When deriving the second equality, 
we have used (\ref{eq:mutor}), 
(\ref{eq:relationamongrandrs}) 
and (\ref{eq:adscftdictionary}). 
From (\ref{eq:instantonactionntom}) together with 
the non-negativity of the Euclidean action (\ref{eq:positivity}), 
one finds the condition for the partition of $J$:
\[
\sum^n_{i=1} J'^2_i \le \sum^m_{j=1} J^2_j\ . 
\label{eq:conditionforpartition}
\] 
Since this condition always holds if $m \le n$, 
we mostly focus on splitting interactions by setting $m\le n$ unless otherwise stated. 
Joining interactions, \textit{i.e.} $m \ge n$, can be obtained via anti-instantons. 
Note that the condition (\ref{eq:conditionforpartition}) is a necessary condition 
for instantons to exist and the necessary and sufficient condition will be discussed 
in the end of Section \ref{sec:4dlaplaceequationinriemannspaces}.

In the dual CFT it is expected that this type of giant graviton interactions corresponds to 
$(m+n)$-point functions of antisymmetric Schur operators (for sphere giants) 
and symmetric Schur operators (for $AdS$ giants)
 \cite{Corley:2001zk,Balasubramanian:2001nh}.
In fact, the pp-wave limit of 3pt functions 
of (anti-)symmetric Schur operators 
has been discussed in \cite{Takayanagi:2002nv}: 
\[
\langle O^{S^5}_{J} \bar O^{S^5}_{J_1} \bar O^{S^5}_{J_2} \rangle 
&=\sqrt{\frac{(N-J_1)!(N-J_2)!}{(N-J)! N!}} 
\cong e^{-\frac{J_1J_2}{2N}}\ , 
\label{eq:3pointantisymmetricschurs} \\
\langle O^{AdS_5}_{J} \bar O^{AdS_5}_{J_1} \bar O^{AdS_5}_{J_2} \rangle 
&= \sqrt{\frac{(N+J-1)!(N-1)!}{(N+J_1-1)!(N+J_2-1)!}} 
\cong e^{\frac{J_1J_2}{2N}}\ ,
\label{eq:3pointsymmetricschurs}
\]   
where $O^{S^5}_{J}$ and $O^{AdS_5}_{J}$ are antisymmetric 
and symmetric Schur operators, respectively. 
These correspond to the $2$-to-$1$ process; two giants with $\mathbb{J}_1 \oplus \mathbb{J}_2$ at $t=-\infty$ 
joining into one giant with $\mathbb{J}$ at $t=+\infty$.  

We thus find the exact agreement within our approximation between the 3pt function 
of antisymmetric Schur operators (\ref{eq:3pointantisymmetricschurs}) 
and the instanton amplitude, since we found
\[
e^{-S_E} = e^{-\frac{J_1J_2}{2N}}\ , 
\label{eq:3s5giants}
\]  
for $J'=J_1+J_2$ in (\ref{eq:instantonactionntom}). 
Note that this is exponentially small in the range $N^{1/2}\ll J\ll N$  but remains finite at large $N$. 
The 3pt function of symmetric Schur operators (\ref{eq:3pointsymmetricschurs}), however,  
cannot correspond to instantons since it grows exponentially as opposed to damping, whereas the instanton action was proven to be always positive. 
We will not resolve this puzzle concerning $AdS$ giants raised in \cite{Takayanagi:2002nv}
and only focus on interactions of sphere giants in the rest of our paper. 

As we will show later, this agreement for antisymmetric Schur operators persists to generic $(m+n)$-point functions, {\it i.e.} to the instantons   
interpolating $m$ sphere giants at $t=-\infty$ 
and $n$ sphere giants at $t=+\infty$.

\section{Four-dimensional Laplace equation in Riemann spaces}
\label{sec:4dlaplaceequationinriemannspaces}
We wish to find solutions to the instanton equation (\ref{eq:instantoneqi}) 
when the matrix size $J$ is very large. 
In the case of the BMN matrix model, the instanton equation analogous 
to (\ref{eq:instantoneqi}) can be mapped to the 3d Laplace equation and various solutions, such as  
one membrane splitting into two membranes, have been found  \cite{Kovacs:2015xha}. 
In this section we show that the instanton equation (\ref{eq:instantoneqi}) can be mapped to 
the 4d Laplace equation following the procedure laid out in \cite{Kovacs:2015xha}. 
As will be shown later, the key observation in \cite{Kovacs:2015xha} is the following relations, as illustrated in Fig. \ref{fig:mton}:
\[
 [\#\,\,\text{of giants at $t=-\infty$}] &\,\,=\,\, [\#\,\,\text{of point charges}]\ , \\
[\#\,\,\text{of giants at $t=+\infty$}] &\,\,=\,\, [\#\,\,\text{of sheets of Riemann space}]\ . 
\label{eq:dictionaryggint}
\]   
We begin with making a change of variables, 
\[
X^i(t) = \sqrt{ \frac{2\mu}{RT}} \ e^{-\mu t} Z^i(s)\ , \ \ \ 
s=e^{-2\mu t}\ , 
\label{eq:changevariables} 
\]
and the instanton equation (\ref{eq:instantoneqi}) 
can be rewritten in terms of the new variables,  
\[
\frac{\text{d} Z^i}{\text{d} s} 
=  \frac{2\pi^2}{3!} \epsilon^{ijkl} 
[Z^j,Z^k,Z^l,\Upsilon_5]\ .
\label{eq:basuharvey}
\]
We note that this  is mathematically the same as the Basu-Harvey equation \cite{Basu:2004ed} 
which describes M$2$ branes ending on M$5$ branes. 
This connection to the Basu-Harvey equation will be exploited in the later section. 

In order to find the solutions describing giant graviton interactions, they have to asymptote to the vacua (static giant gravitons) at the infinite past and future:\footnote{These boundary conditions can be shifted by identify matrices $Z^i (s)\to Z^i (s)-a^i \mathbb{I}_{J\times J}$.}   
\[
Z^i (s) &\cong \sqrt{\frac{RT}{2\mu s}} X^i_0 (-\infty) + \cdots\ , 
\ \ \ \text{for} \ \ \ s \to \infty\ ,
\label{eq:boundaryconditionbasuharvey1} \\
Z^i (s) &\cong \sqrt{\frac{RT}{2\mu s}} X^i_0 (\infty) + \cdots\ , 
\ \ \ \text{for} \ \ \ s \to 0\ ,
\label{eq:boundaryconditionbasuharvey2}
\]
where the ellipses indicate subleading terms and 
$X^i_0(\pm \infty)$ are $J \times J$ representations of $Spin (4)$ satisfying (\ref{eq:adsgiant}) corresponding to the clusters of giants. 
$X^i_0(\pm \infty)$ also need to satisfy the necessary and sufficient condition for the existence of instantons 
discussed in the end of Section \ref{sec:4dlaplaceequationinriemannspaces}.    
These set the boundary conditions for the solutions we are after.

When the matrix size is very large,  the matrices $Z^i$ can be approximated by 
the functions $z^i(s,\sigma^{\mu})$ 
and the quantum Nambu 4-bracket $[\ast,\ast,\ast,\Upsilon_5]$ 
by the Nambu 3-bracket. 
This is the \lq\lq classicalisation''  of 
the brackets, reversing the procedure (\ref{eq:matrixregularisation1}) - (\ref{eq:matrixregularisation4}). 
Then the Basu-Harvey equation (\ref{eq:basuharvey}) can be 
approximated by 
\[
\frac{\partial z^i}{\partial s} 
= 
- \frac{[\sigma]}{3! J} 
\epsilon^{ijkl} 
\{ 
z^j,z^k,z^l
\}
= - \frac{[\sigma]}{J}\epsilon^{ijkl} \frac{\partial z^j}{\partial \sigma^1}\frac{\partial z^k}{\partial \sigma^2}\frac{\partial z^l}{\partial \sigma^3}\ ,
\label{eq:continuumbasuharvey}
\] 
which can be locally mapped to the 4d Laplace equation 
as shown in Appendix \ref{sec:ap:derivationoflaplaceequation}. 
Essentially, this map can be made by interchanging the role of 
dependent and independent variables:
\[
(z^1,z^2,z^3,z^4) \ \ \ 
\leftrightarrow
\ \ \ 
(s,\sigma^1,\sigma^2,\sigma^3)\ .  
\label{eq:hodographtr}
\]  
This means solving $s$ as a function of $z^i$:
\[
s = \phi (z^i)\ . 
\label{eq:sisphi}
\]
Using this hodograph transformation 
the equation (\ref{eq:continuumbasuharvey}) 
is mapped to the 4d Laplace equation 
(see Appendix \ref{sec:ap:derivationoflaplaceequation} for details): 
\[
\sum^{4}_{i=1}
\left( 
\frac{\partial}{\partial z^i} 
\right)^2 \phi = 0\ .  
\label{eq:laplaceequation}
\]
We will then find solutions to the Laplace equation (\ref{eq:laplaceequation}) 
corresponding to splitting interactions of concentric giants. 
The equipotential surface provides the profile of giant gravitons for a given $s$ 
in the $z$-space.

Let us see how a single fuzzy three-sphere can be described by 
a solution to the Laplace equation. 
A single fuzzy $S^3$ corresponds to the $J \times J$ irreducible representation of $Spin(4)$ 
which is a static solution to the instanton equation (\ref{eq:adsgiant}) and
denoted by the matrices $X^i_0$. 
By the change of variables (\ref{eq:changevariables}) 
we can map $X^i_0$ to the matrices $Z^i_0$ representing the spatial coordinates of giants:
\[
Z^i_0 = \sqrt{\frac{2\mu}{RTs}} X^i_0\ . 
\label{eq:z0}
\] 
When the matrix size $J$ is very large, 
we replace the matrices $X^i_0$ and $Z^i_0$, 
by functions $x^i$ and $z^i$, 
and accordingly, (\ref{eq:z0}) is approximated by 
\[
z^i = \sqrt{\frac{2\mu}{RTs}} x^i\ , 
\label{eq:contz0}
\] 
where $x^i$ form a three-sphere of radius (\ref{eq:radiusofgiant}): 
\[
x^i = r n^i\ , 
\ \ \ \sum^{4}_{i=1} (x^i)^2 = r^2\ . 
\label{eq:singlethreesphere} 
\]
Here $n^i$ is the unit vector normal to the three-sphere 
(see Appendix \ref{sec:ap:threespheres} for details). 
One can solve $s$ as a function of $z^i$ by (\ref{eq:contz0}): 
\[
s = \frac{J}{4\pi^2 |z^i|^2}\ . 
\label{eq:coulombpotential}
\]    
This is nothing but the Coulomb potential in four dimensions 
with charge $J$ at the origin, 
which, of course, solves the Laplace equation (\ref{eq:laplaceequation}).  
Through this simple example, we have learned that 
a single giant graviton with angular momentum $J$ 
can be described by the 4d Coulomb potential for point charge $J$. 

We shall generalise this to the instantons interpolating between
$m$ (concentric) giants at $t=-\infty$ and 
$n$ (concentric) giants at $t=+\infty$. 
As will be explained in Section \ref{sec:splittinginteractionsofgiantgravitons}, 
these splitting processes of concentric giants 
are described by the solutions to the 4d Laplace equation in multi-sheeted Riemann spaces 
rather than the ordinary 4d Euclidean space $\mathbb{R}^4$.  
The 4d Riemann spaces are a four-dimensional analogue of  
2d Riemann surfaces, and the precise definition will be given 
in \ref{sec:hypertoroidalcoordinatesandriemannspaces}.

The use of Riemann spaces has been first emphasised  
in the study of membrane interactions \cite{Kovacs:2015xha}: 
They considered splitting interactions of concentric spherical membranes 
with large angular momenta as instantons in the BMN matrix model.    
It was found that the instanton equation in their case can be locally mapped to 
the $3$d Laplace equation and the splitting interactions correspond to the Coulomb potentials in multi-sheeted 3d Riemann spaces.

\subsection{Hypertoroidal coordinates and Riemann spaces}
\label{sec:hypertoroidalcoordinatesandriemannspaces}

We introduce the coordinates which are particularly useful 
for studying the solutions to the $4$d Laplace equation in multi-sheeted Riemann spaces. 
In this paper we call them the hypertoroidal coordinates.

\begin{figure}[h]
\centering
\includegraphics[width=3.5in]{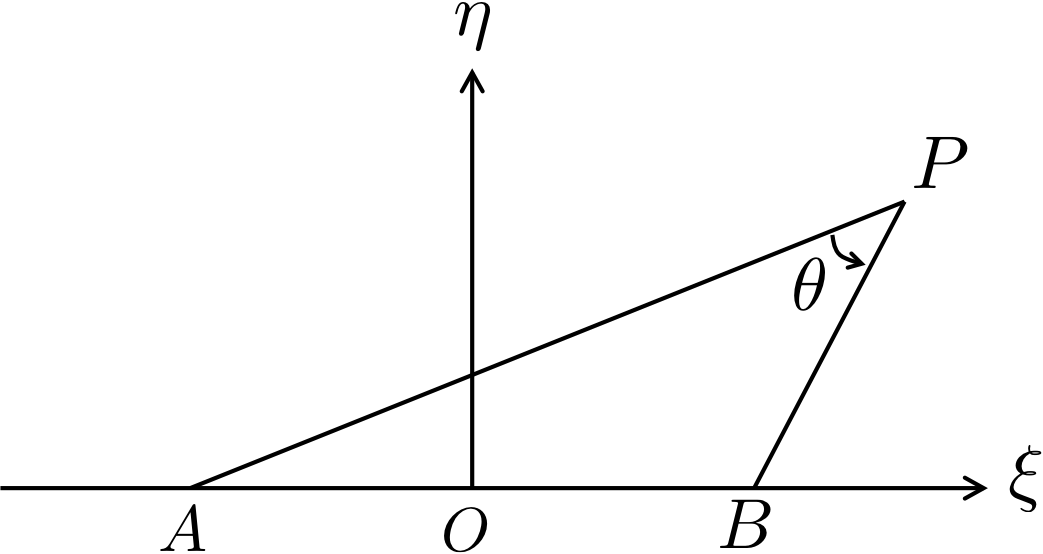}
\caption{Bipolar coordinates $(\rho,\theta)$: 
$\theta = \angle APB$, 
$\rho = \log |AP|/|BP|$ and 
$|AO|=|BO|=a$.
}
\label{fig:bipolar}
\end{figure}

To set up, we consider a point $P$ designated by $(\rho,\theta)$ in the bipolar coordinates   
relating to the two-dimensional Cartesian coordinates $(\xi,\eta)$ as 
\[
\xi={a\sinh\rho\over\cosh\rho-\cos\theta}\ , \ \ \ 
\eta={a\sin\theta\over\cosh\rho-\cos\theta}\ . 
\label{eq:bipolar}
\]
The definition of $\rho$ and $\theta$ is given as follows. 
We call two points in the two-dimensional Cartesian coordinates, 
$(-a,0)$ and $(a,0)$, $A$ and $B$, respectively (see Fig. \ref{fig:bipolar}). 
The angle $\angle{APB}$ is denoted by $\theta$ defined to be in the interval $[-\pi,\pi]$;
\[
\rho = \log \frac{|AP|}{|BP|}\ ,
\] 
where $|AP|$ and $|BP|$ are 
the lengths of segments, $AP$ and $BP$, respectively and 
by definition $\rho \in (-\infty,\infty)$.  

If we extend the interval of $\theta$ from $[-\pi,\pi]$ to $[-\pi,3\pi]$,   
the bipolar coordinates become multi-valued. 
To make the coordinates single-valued, 
we introduce a cut, say the segment $AB$, 
and stitch two copies of $\mathbb{R}^2$'s by the cut $AB$ 
such that if $\theta \in [-\pi,\pi]$, the space belongs to an $\mathbb{R}^2$ 
and if $\theta \in [\pi, 3\pi]$, it does to the other $\mathbb{R}^2$. 
This space is a $2$d (two-sheeted) Riemann space, 
which can be easily extended to an $(n+1)$-sheeted Riemann space 
if one considers 
the interval of $\theta$ to be $[-\pi,\pi +2\pi n]$ with $n$ being positive integer.  
In that case we prepare $(n+1)$ copies of $\mathbb{R}^2$ 
such that each $\mathbb{R}^2$ is specified 
by the different interval of $\theta$, $[-\pi + 2\pi m,  \pi + 2\pi m]$ with $m=0,1,\cdots,n$. 
We then stitch the $(n+1)$ $\mathbb{R}^2$'s together by the cut $AB$, 
resulting in an $(n+1)$-sheeted Riemann space.

We introduce the hypertoroidal coordinates as a $4$d extension of the bipolar coordinates.\footnote{A $3$d extension of the bipolar coordinates is called the toroidal coordinates or the peripolar coordinates.}    
This can be constructed by rewriting the $4$d spherical coordinates
\[
(z^1,z^2,z^3,z^4) 
= 
r(\cos \lambda, \sin \lambda \cos \varphi, \sin \lambda \sin \varphi \cos \omega, \sin \lambda \sin \varphi \sin \omega)\ ,  
\label{eq:sphericalcoordinates}
\] 
as 
\[
(z^1,z^2,z^3,z^4) 
= (\eta,\xi \cos \varphi, \xi \sin \varphi \cos \omega,\xi \sin \varphi \sin \omega)\ ,
\label{hypertoroidal1}
\]
where 
\[
r\cos\lambda=\eta={a\sin\theta\over\cosh\rho-\cos\theta}\ ,\qquad
r\sin\lambda=\xi={a\sinh\rho\over\cosh\rho-\cos\theta}\ .
\label{hypertoroidal2}
\]
\begin{figure}[h]
\centering
\includegraphics[width=4in]{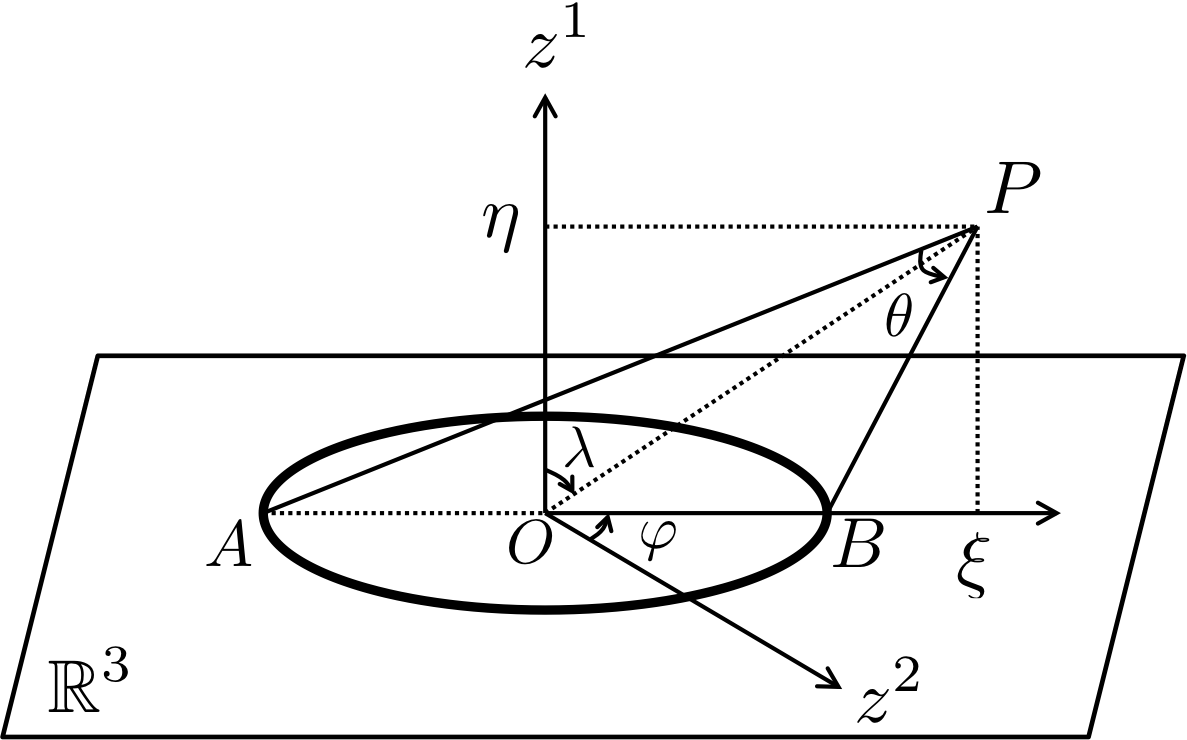}
\caption{Hypertoroidal coordinates $(\rho,\theta,\varphi,\omega)$:
The region inside the circle drown by the heavy line stands for 
a three-ball of radius $a$ embedded in $\mathbb{R}^3$.
This three-ball plays a role 
analogous to a branch cut in a $2$d Riemann space 
once we extend the interval of $\theta$.}
\label{fig:hypertoroidal}
\end{figure}
The Fig. \ref{fig:hypertoroidal} is a graphical expression of the hypertoroidal coordinates 
in which $r = |OP|$.  
Since $\xi$ and $\eta$ are the same as (\ref{eq:bipolar}), 
the interval of $\rho$ and $\theta$ is $\theta \in [-\pi,\pi]$ and $\rho \in (-\infty,\infty)$, respectively. 
The angles, $\varphi$ and $\omega$, are respectively defined to be in the intervals, $[0,\pi]$ and $[0,2\pi]$.

Extending the interval of $\theta$ from $[-\pi,\pi]$ to $[-\pi,\pi+2\pi n]$ 
with $n$ being positive integer  
as in the case of the bipolar coordinates,  
the hypertoroidal coordinates become multi-valued.   
In order to make the coordinates single-valued, 
we need to introduce an object analogous to a cut in 
a $2$d Riemann space 
which is a three-ball of radius $a$ 
located at $\theta = -\pi + 2\pi m$ 
with $m=0,1, \cdots, n$ (see Fig. \ref{fig:hypertoroidal}). 
We call this three-ball a branch three-ball. 
As before we prepare $(n+1)$ copies of $\mathbb{R}^4$ such that 
each $\mathbb{R}^4$ is designated by 
the different interval of $\theta$, $[-\pi + 2\pi m, \pi+ 2\pi m]$ 
with $m = 0,1, \cdots, n$. 
Gluing the $(n+1)$ $\mathbb{R}^4$'s at the branch three-ball, 
we can construct a $4$d $(n+1)$-sheeted Riemann space.

It goes back to $1896$ when Sommerfeld first considered the three-dimensional Laplace equation 
in Riemann spaces \cite{Sommerfeld:1896, Sommerfeld:1899}.  We shall extend his idea to the four-dimensional space
for the purpose of finding solutions describing splitting interactions of concentric giant gravitons, following the success of  
\cite{Kovacs:2015xha} in their application of \cite{Davis:1971, Hobson:1900} to membrane interactions.

\subsection{Splitting interactions of giant gravitons}
\label{sec:splittinginteractionsofgiantgravitons}

We now discuss in detail the construction of solutions to the 4d Laplace equation in Riemann spaces which describe 
splitting interactions of concentric giant gravitons with large angular momenta. 
As we have seen in Section \ref{sec:4dlaplaceequationinriemannspaces}, 
after the map to the Laplace equation, 
the snapshots of giant gravitons at time $s$ in the $z$-space are the equipotential surfaces $s=\phi (z^i)$, and 
a single giant with angular momentum $J$ corresponds to the Coulomb potential created by a point charge $J$.  
From (\ref{eq:changevariables}) the infinite past and future  
correspond to $s=+\infty$ and $s=0$, respectively.  

\begin{figure}[h]
\centering
\includegraphics[width=4.4in]{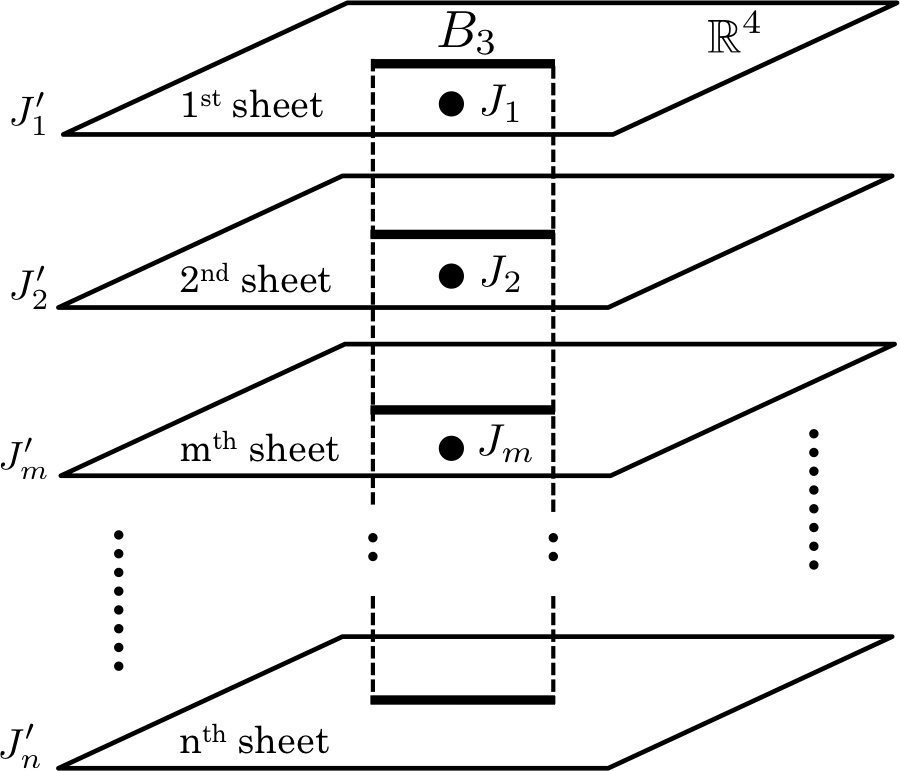}
\caption{$m$ point charges in an $n$-sheeted Riemann space: 
The thick line segments and blobs are the branch three-balls and point electric charges, respectively. The branch three-balls are all identified.
The number of point charges $m$ corresponds to the number of giant gravitons at $t=-\infty$ 
and the number of sheets of the Riemann space $n$ to the number of giant gravitons at $t=+\infty$.}
\label{fig:mton}
\end{figure}

The construction of our solutions goes as follows. (See Fig. \ref{fig:mton}):
In a $4$d Riemann space with  $n$-sheets we place 
$m$ point charges but only allow at most one charge per a single sheet.
This corresponds to the instanton interpolating one vacuum
$\mathbb{J}_1 \oplus \mathbb{J}_2 \oplus \cdots \oplus \mathbb{J}_m$ at $t=-\infty$ 
and the other
$\mathbb{J}'_1 \oplus \mathbb{J}'_2 \oplus \cdots \oplus \mathbb{J}'_n$ at $t=\infty$ 
with the constraints $J_1 + \cdots +J_m =J'_1 + \cdots +J'_n =J$ and $m \le n$.   
Simply put, the correspondence is 
\[
 [\#\,\,\text{of giants at $t=-\infty$}] &\,\,=\,\, [\#\,\,\text{of point charges}]\ , \\
[\#\,\,\text{of giants at $t=+\infty$}] &\,\,=\,\, [\#\,\,\text{of sheets of Riemann space}]\ . 
\]   
The number of point charges equals the number of giants at $t=-\infty$, and 
the number of sheets is the number of giants at $t=+\infty$. 
This is because 
the infinite past $s=+\infty$ corresponds to the diverging potential at the locations of $m$ point charges
and the infinite future $s=0$ to the asymptotic infinities, $z^i\to\infty$, in the Riemann space.
The electric flux runs through the branch three-ball to different sheets and escapes to the asymptotic infinities. 
By construction the necessary condition (\ref{eq:conditionforpartition}), 
or equivalently, the condition $m\le n$ 
is automatically satisfied.   
This construction is the 4d analog of the one for membrane interactions \cite{Kovacs:2015xha}. 

The above construction can be worked out explicitly by applying Sommerfeld's extended image technique \cite{Sommerfeld:1896, Sommerfeld:1899}: 
To begin with,  we consider the $4$d Coulomb potential
 \[
\phi (z^i) 
= \frac{J}{4\pi^2 |z^i-z^i_0|^2} 
=:\frac{J}{4\pi^2 \mathcal{R}^2}\ ,
\label{eq:4dcoulombpotential}
\] 
where $J$ is a point charge placed at $z^i=z^i_0$ in $\mathbb{R}^4$. Using the hypertoroidal coordinates $(\rho,\theta,\varphi,\omega)$ 
defined in (\ref{hypertoroidal1}) and (\ref{hypertoroidal2}),  
the distance squared from the charge is expressed as 
\[
\mathcal{R}^2  
= 
\frac{2a^2 ( \cosh \alpha - \cos (\theta - \theta_0) )}{(\cosh \rho - \cos \theta)(\cosh \rho_0 - \cos \theta_0)}\ , 
\label{eq:rsquared}
\] 
where we defined
\[
\cosh \alpha := 
\cosh \rho \cosh \rho_0 
- \sinh \rho \sinh \rho_0 
(
\cos \varphi \cos \varphi_0 
+ \sin \varphi \sin \varphi_0 
\cos (\omega - \omega_0)
)\ . 
\label{eq:coshalpha}
\] 
As explained in Section \ref{sec:hypertoroidalcoordinatesandriemannspaces}, 
once we extend the interval of the angle $\theta$ from $[-\pi,\pi]$ to $[-\pi,\pi+2\pi (n-1)]$ 
with $n \ge 2$,  
the hypertoroidal coordinates become multi-valued and 
we can construct an $n$-sheeted Riemann space by 
stitching $n$ $\mathbb{R}^4$'s at the branch three-ball of radius $a$ 
and make the coordinates single-valued. 
Because the distance squared $\mathcal{R}^2$ in \eqref{eq:rsquared}  is periodic in the angle $\theta$, so is the Coulomb potential (\ref{eq:4dcoulombpotential}) and there must be charges placed in every single sheet at the same location.
In other words, the Coulomb potential is an $n$-charge solution where every single sheet has one charge at the same location in $\mathbb{R}^4$.

In order to find generic $m\le n$ charge solutions,  we first look for the electrostatic potential created by a single charge placed in only one of the $n$ sheets in the Riemann space. This is going to serve as the building block for the construction of more general potentials.
One can distill a single charge contribution from the Coulomb potential (\ref{eq:4dcoulombpotential}) by expressing it 
as a contour integral and deforming the contour \cite{Sommerfeld:1896, Sommerfeld:1899}.

\subsubsection{Coulomb potential in two-sheeted Riemann space}
\label{sec:coulombpotentialintwosheetedriemannspace}

Let us first consider the two-sheeted case. 
We complexify the angle $\theta$ and introduce the complex variable $\zeta=e^{i\theta/2}$ 
which covers the two-sheeted Riemann space.  
The Coulomb potential (\ref{eq:4dcoulombpotential}) can then be expressed as a contour integral
\[
\phi(z^i)&={J\over 8\pi^3i}\oint_{C_{\theta}}\text{d}\zeta'{\mathcal{R}^{-2}(e^{i\theta'}\to\zeta'^2)\over\zeta'-e^{i\theta/2}}
{\cosh\rho-\cos\theta\over\cosh\rho-\cos\theta'}\nn\\
&={J\over 16\pi^3}\oint_{C_{\theta}} \text{d}\theta'{\mathcal{R}^{-2}(e^{i\theta'}\to\zeta'^2)\over 1-e^{i(\theta-\theta')/2}}
{\cosh\rho-\cos\theta\over\cosh\rho-\cos\theta'} \notag \\
&={J\over 32\pi^3a^2}\oint_{C_{\theta}} \text{d}\theta'
{(\cosh\rho_0-\cos\theta_0)(\cosh\rho-\cos\theta)\over \left(1-e^{i(\theta-\theta')/2}\right)\left(\cosh\alpha-\cos(\theta'-\theta_0)\right)}\ , 
\label{eq:twosheetscontourintegral1}
\]
where the contour $C_{\theta}$ is a unit circle surrounding $\zeta=e^{i\theta/2}$. 
The factor $(\cosh\rho-\cos\theta) / (\cosh\rho-\cos\theta')$ 
in the integrand is inserted to ensure that the integrand vanishes at $\theta'=\pm i\infty$. 

Besides the poles at $\theta'=\theta+4k\pi$ with $k\in\mathbb{Z}$, the integrand in (\ref{eq:twosheetscontourintegral1}) has the poles at 
\[
\theta'=\theta_0+2k\pi\pm i\alpha\ .
\label{eq:twosheetedpoles}
\]  
\begin{figure}[h]
\centering
\includegraphics[width=5in]{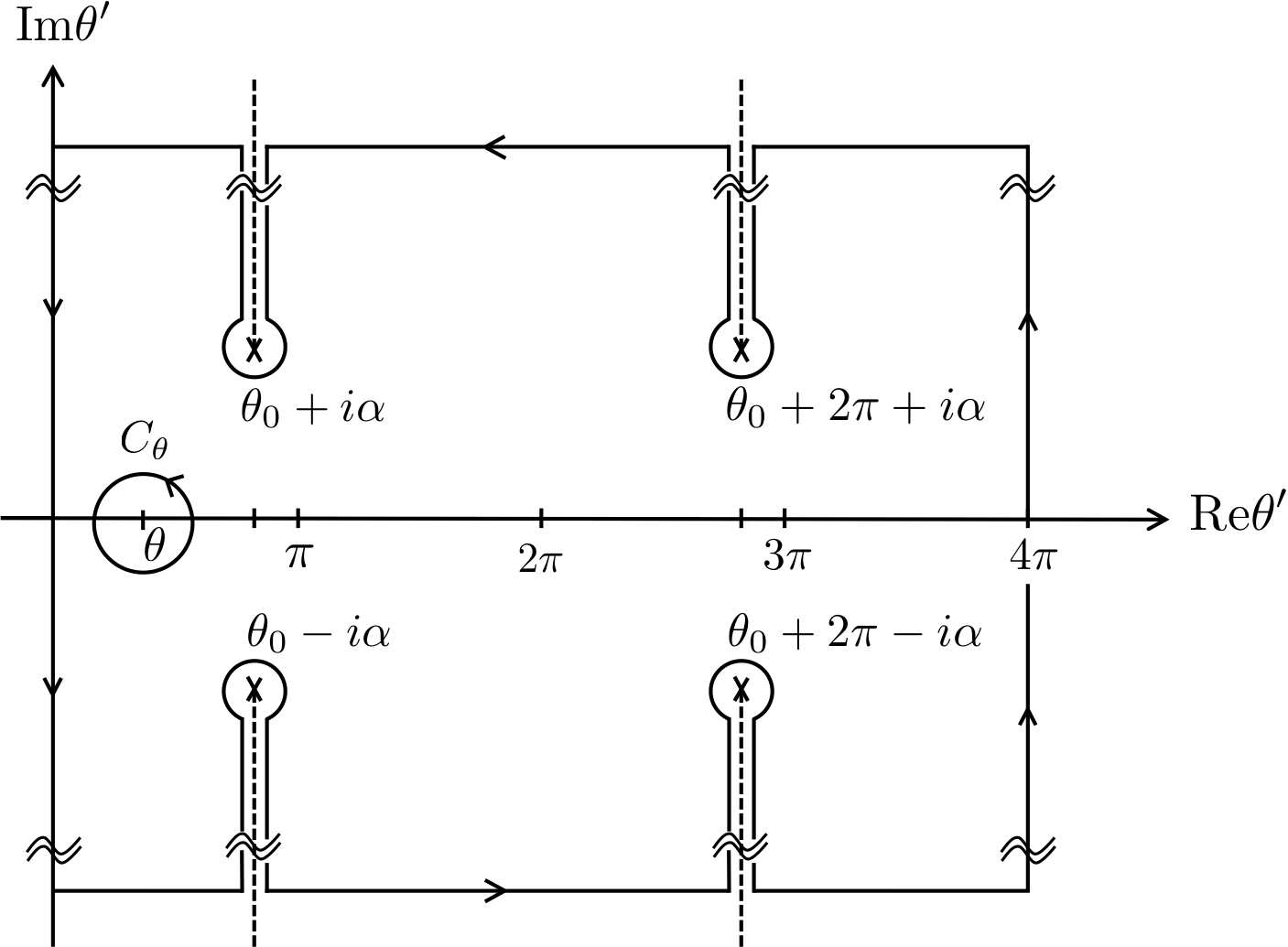}
\caption{The contour deformation}
\label{fig:contours}
\end{figure}  
We now deform the contour $C_{\theta}$ to a rectangle of width $4\pi$ 
and an infinite height while avoiding the poles at $\theta'=\theta_0\pm i\alpha$ and $\theta_0+2\pi\pm i\alpha$ 
(see Fig. \ref{fig:contours}). 
The contributions from the vertical edges cancel out owing to the periodicity, 
and those from the horizontal edges at infinity simply vanish. 
The single charge contribution is extracted 
as the residue of a pole and its pair in the lower-half plane. 
Note first that at $\theta'\simeq \theta_0+2k\pi\pm i\alpha$ we have
\[
\cosh\alpha-\cos(\theta'-\theta_0)\simeq \pm i (\theta'-\theta_0-2k\pi\mp i\alpha)\sinh\alpha\ . 
\label{eq:twosheetsapprox}
\]
The relevant part of the contour comes in from infinity, 
encircles a pole clockwise and goes back to infinity, picking up the residue. 
The single charge potential is thus found to be
\[
\phi_{k=0}(z^i) 
=-{J\over 32\pi^3a^2}\oint_{C_{\theta_0+i\alpha}+C_{\theta_0-i\alpha}} 
\!\!\! \text{d} 
\theta'{(\cosh\rho_0-\cos\theta_0)(\cosh\rho-\cos\theta)\over \left(1-e^{i(\theta-\theta')/2}\right)\left(\cosh\alpha-\cos(\theta'-\theta_0)\right)}\ .
\label{eq:singlechargepot}
\]
The consistency requires 
\[
\phi(z^i)=\phi_{k=0}(z^i)+\phi_{k=1}(z^i)\ ,
\label{eq:2chargesum}
\]
where the second term is the contribution from a charge on the second sheet.
Carrying out the contour integral (\ref{eq:singlechargepot}), 
we find that 
\[
\phi_{k=0}(z^i)
={J\over 4\pi^2 \mathcal{R}^2}\left(\half+\half{\cos{\theta-\theta_0\over 2}\over \cosh{\alpha\over 2}}\right)\ . 
\label{eq:twosheetssinglecharge}
\]
One can check that (\ref{eq:2chargesum}) holds by noting that $\phi_{k=1}(\theta)=\phi_{k=0}(\theta+2\pi)$.

\subsubsection{Coulomb potential in $n$-sheeted Riemann space}
\label{sec:coulombpotentialinmsheetedriemannspace}

\begin{figure}[htbp]
\vspace{-1cm}
\centering
\includegraphics[height=0.295\textheight]{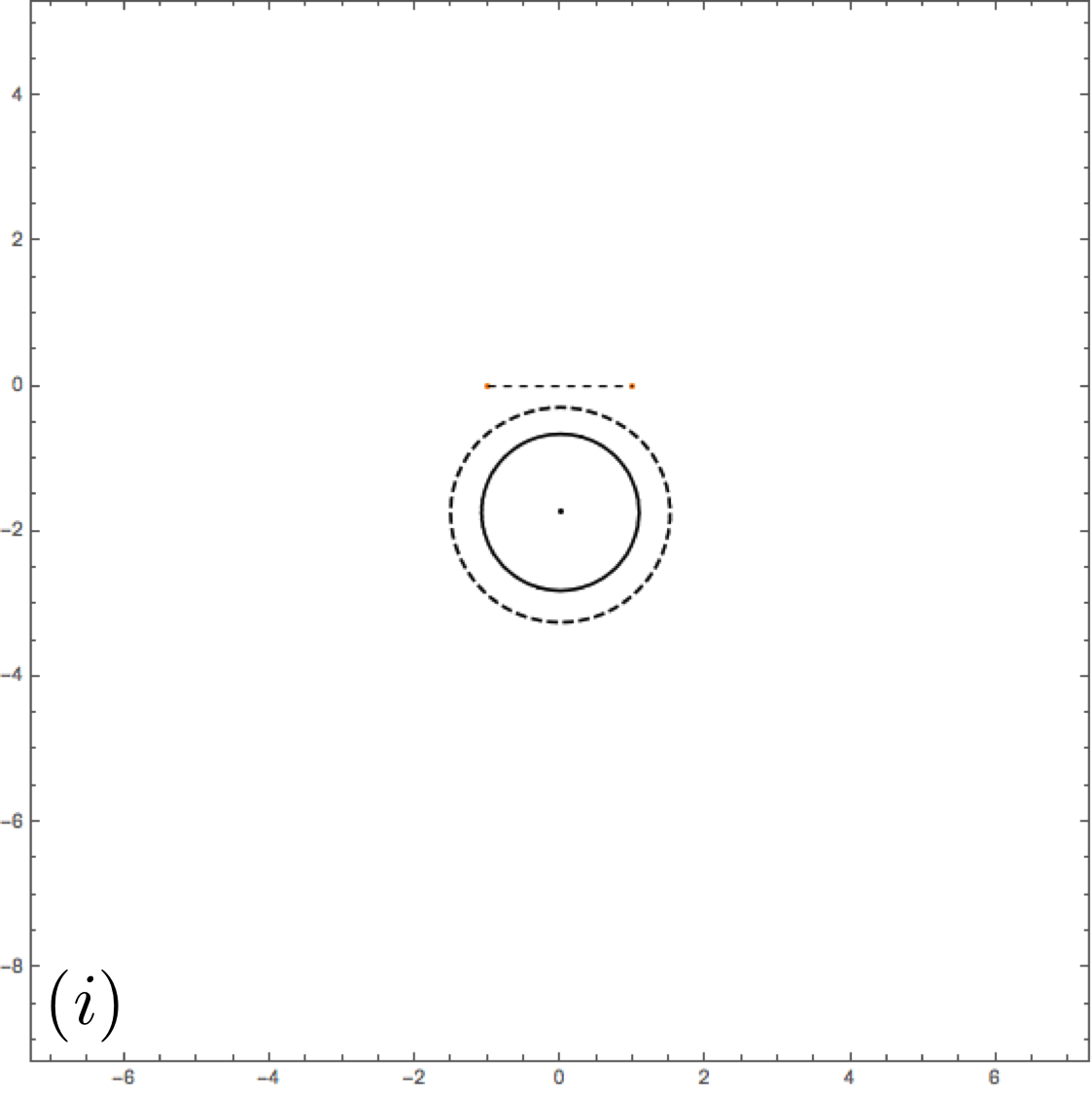} \hspace*{0.2cm}
\includegraphics[height=0.295\textheight]{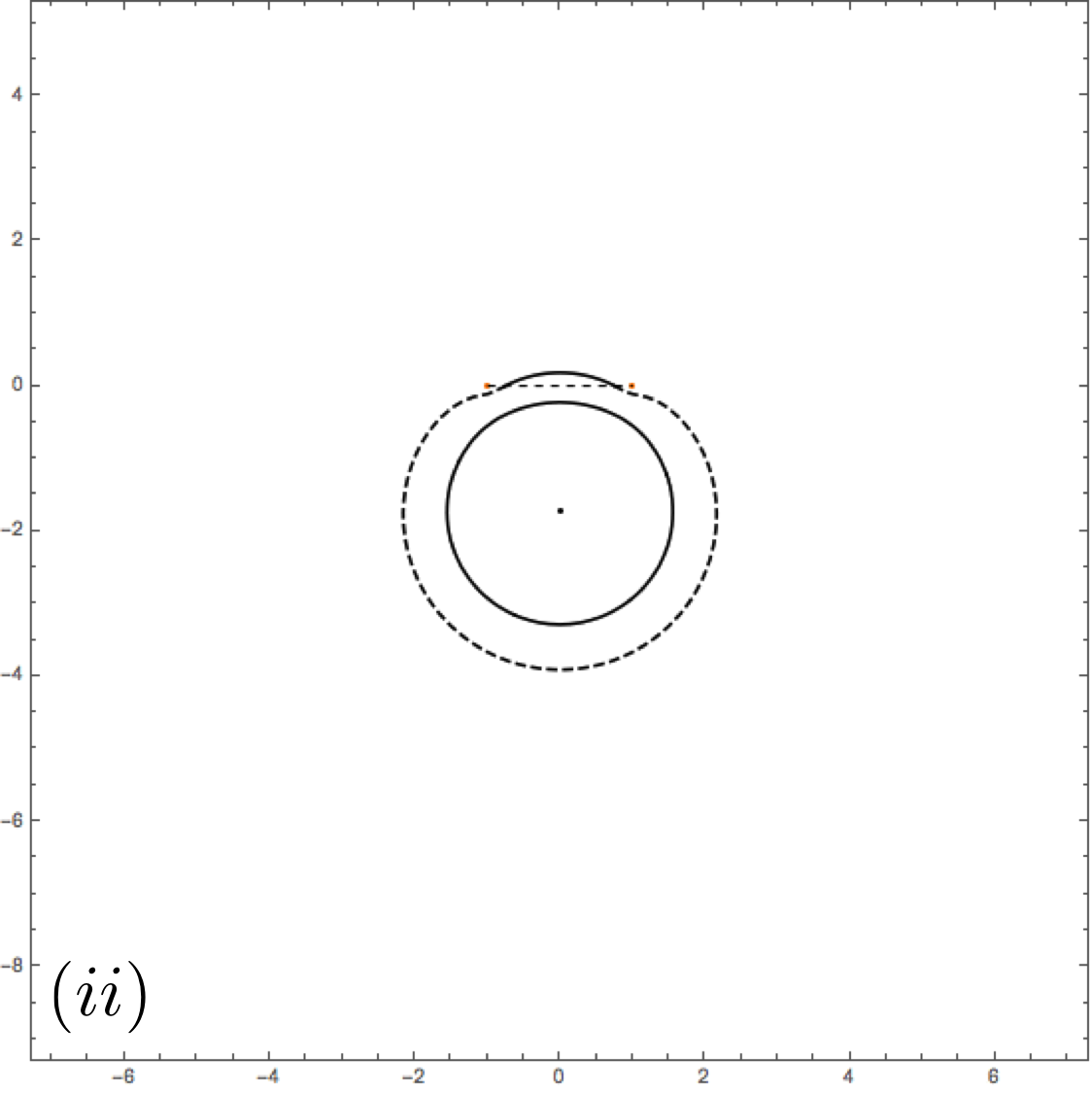} \\
\includegraphics[height=0.295\textheight]{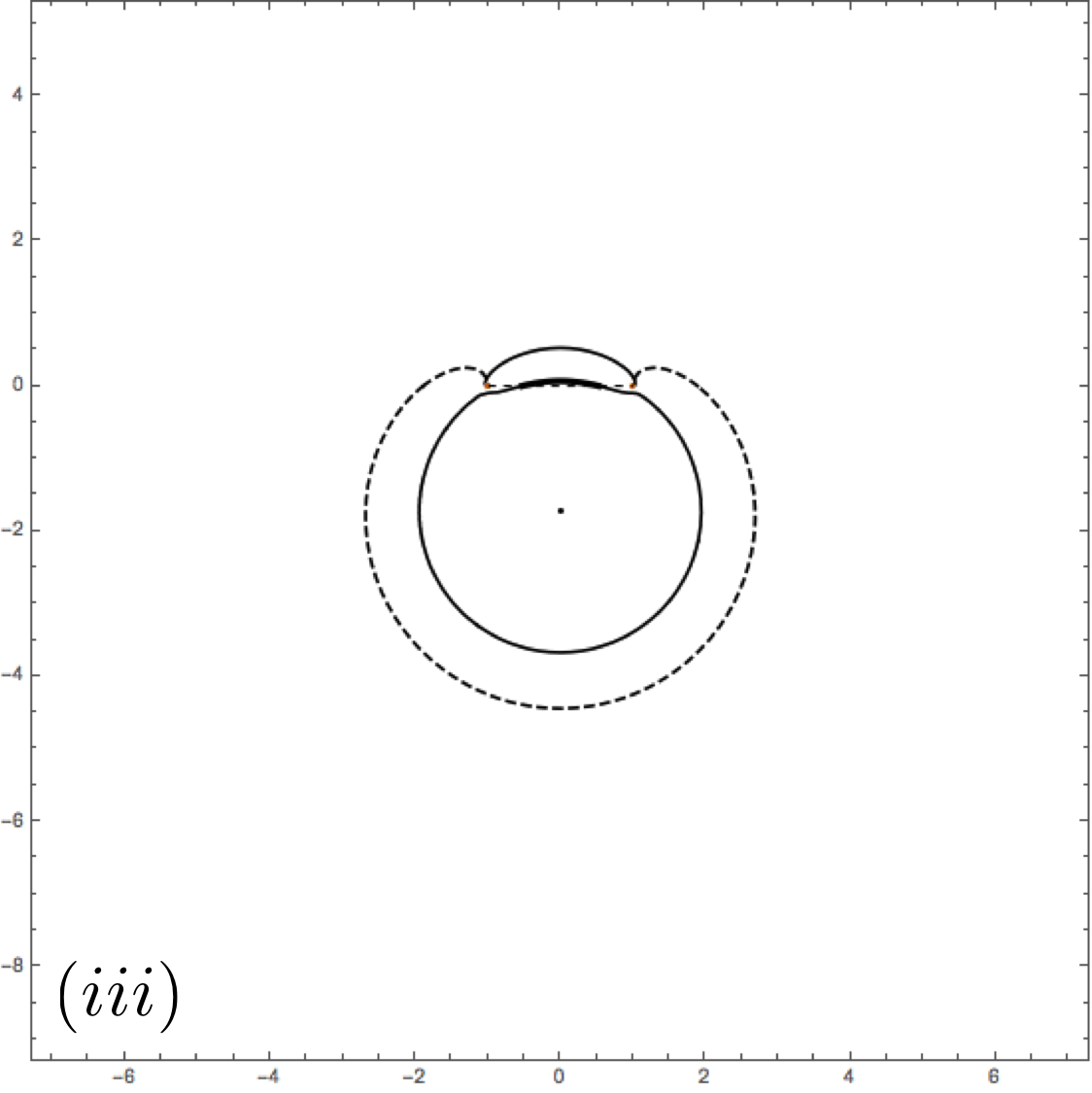} \hspace*{0.2cm}
\includegraphics[height=0.295\textheight]{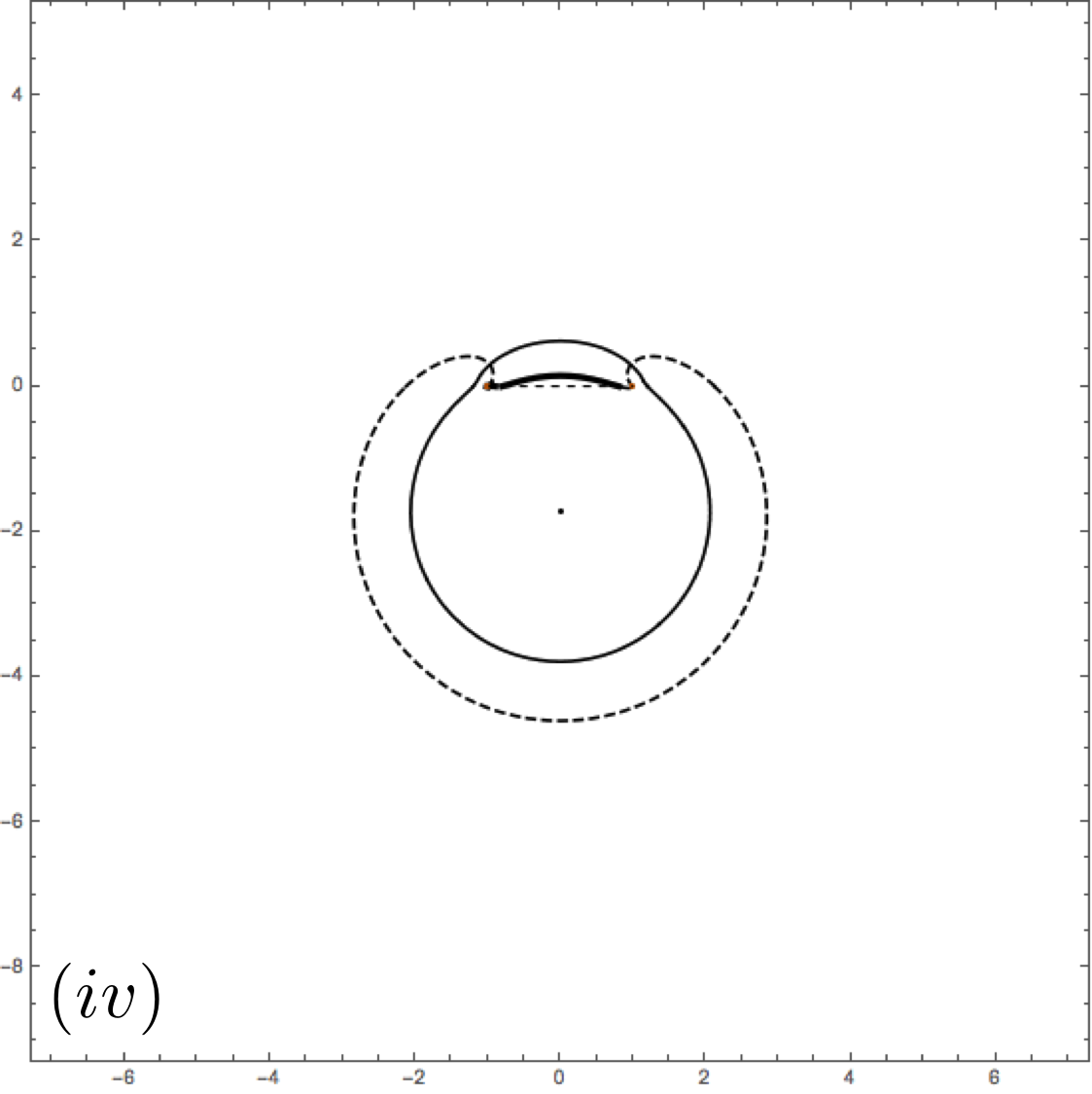} \\
\includegraphics[height=0.295\textheight]{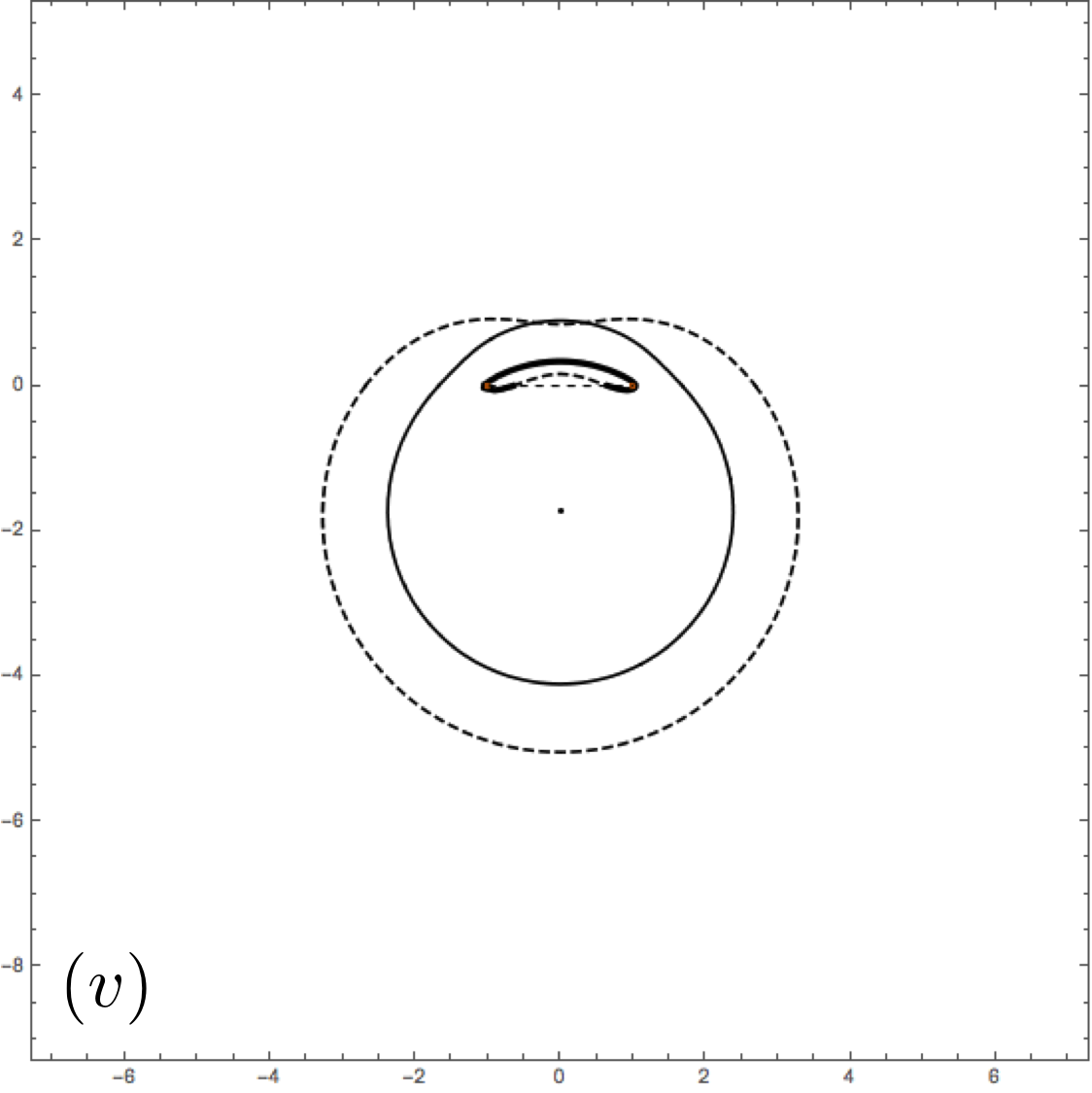} \hspace*{0.2cm}
\includegraphics[height=0.295\textheight]{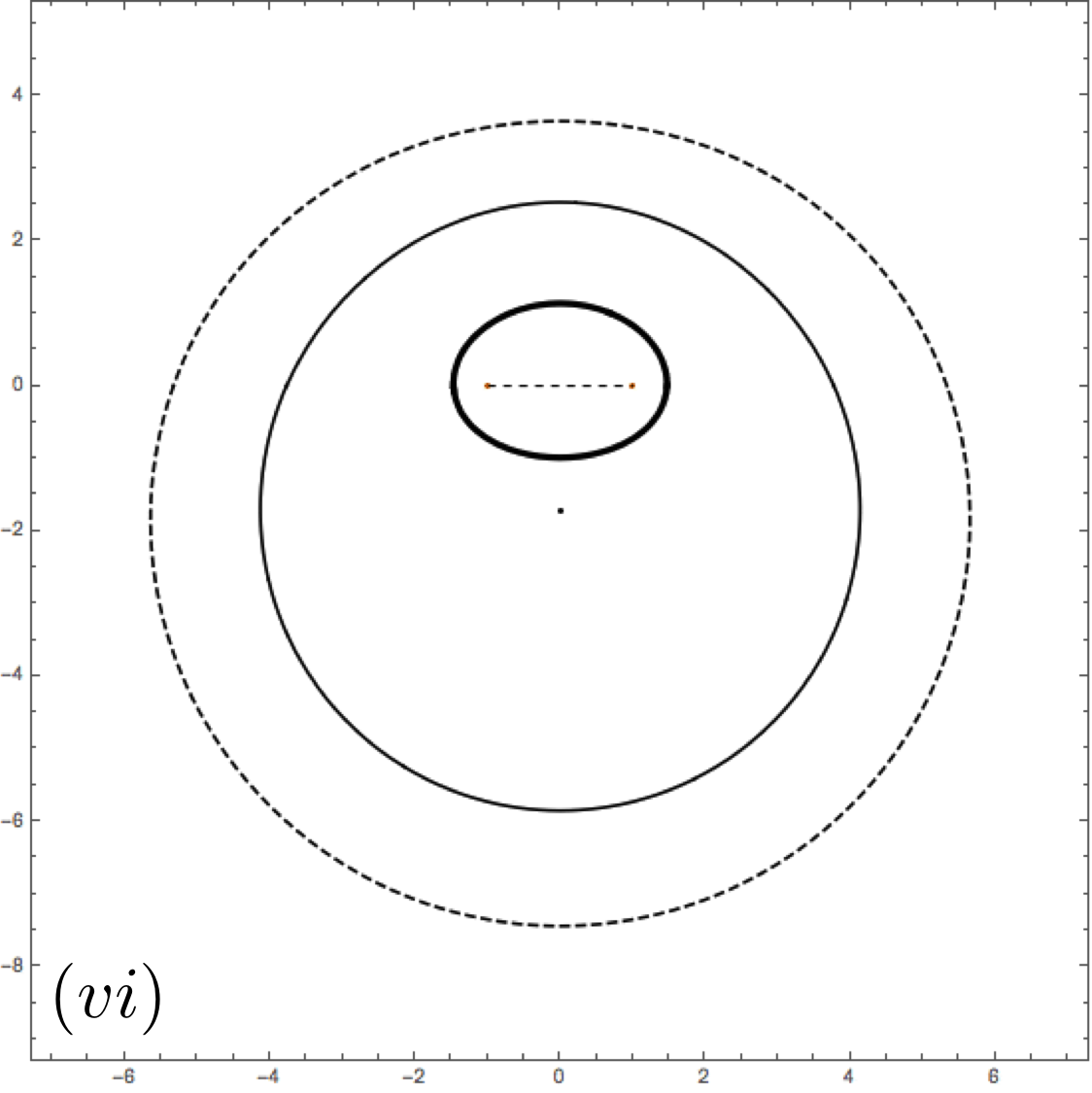} \\
\caption{An evolution of the giant graviton profiles for a process with two initial
concentric giants and three final concentric giants in chronological order:
The thin solid, dashed and thick solid lines indicate (projections of) 
the equipotential surfaces in the 1st-, 2nd- and 3rd-sheets. 
The horizontal dashed segment and the black point are the branch three-ball 
and the point charge, respectively. The splitting interaction is happening from (ii) to (v) through the branch. }
\label{fig:evolutionofgiants}
\end{figure}

It is straightforward to generalise the two-sheet case to the $n$-sheeted Riemann space. 
We start from
\[
\phi(z^i)
={J\over 16n\pi^3a^2}\oint_{C_{\theta}} \text{d}
\theta'{(\cosh\rho_0-\cos\theta_0)(\cosh\rho-\cos\theta)\over \left(1-e^{i(\theta-\theta')/n}\right)\left(\cosh\alpha-\cos(\theta'-\theta_0)\right)}\ ,   
\label{eq:msheetscontourintegral1}
\]
and deform the contour in a similar manner to the two-sheet case. There are poles at $\theta'=\theta+2nk\pi$ with $k\in\mathbb{Z}$ and 
\[
\theta'=\theta_0+2k\pi\pm i\alpha\ . 
\label{eq:msheetedpoles}
\]
Similar to the two-sheet case, 
the rectangle contour with width $2n\pi$ picks up the residues from these poles. 
The Coulomb potential splits into
\[
\phi(z^i)
=\phi_{k=0}(z^i)+\phi_{k=1}(z^i)+\cdots+\phi_{k=n-1}(z^i)\ .
\]
The single charge potential is thus given by
\[
\phi_{k=0}(z^i) 
&=-{J\over 16n\pi^3a^2}\oint_{C_{\theta_0+i\alpha}+C_{\theta_0-i\alpha}} 
\!\!\! \text{d}
\theta'{(\cosh\rho_0-\cos\theta_0)(\cosh\rho-\cos\theta)\over \left(1-e^{i(\theta-\theta')/n}\right)\left(\cosh\alpha-\cos(\theta'-\theta_0)\right)} \notag \\
&={J\over 4\pi^2 \mathcal{R}^2}{\sinh{\alpha\over n}\left(\cosh^2{\alpha\over 2}-\cos^2{\theta-\theta_0\over 2}\right)\over n\sinh{\alpha}\left(\cosh^2{\alpha\over 2n}-\cos^2{\theta-\theta_0\over 2n}\right)} 
=: \phi^{(J)}_n (z^i ; \theta_0)
\ .
\label{eq:msheetsinglechargepot}
\]
By superposing different contributions, it is easy to construct
the solution to the 4d Laplace equation describing general $m$ giants 
$\mathbb{J}_1 \oplus \mathbb{J}_2 \oplus \cdots \oplus \mathbb{J}_m$ 
at $t=-\infty$ splitting into 
$n$ giants 
$\mathbb{J}'_1 \oplus \mathbb{J}'_2 \oplus \cdots \oplus \mathbb{J}'_n$
at $t=\infty$ 
with the constraints $J_1 + \cdots + J_m=J'_1 + \cdots + J'_n=J$ and $m \le n$:  
\[
\phi_{m,n}(z^i) 
= \sum^m_{l=1} 
\phi^{(J_l)}_{n} (z^i; \theta_0 + 2\pi (l-1))\ ,
\label{eq:ntomgiantsolution}
\]
where we defined $\phi^{(J)}_n (z^i ; \theta_0)$ in \eqref{eq:msheetsinglechargepot}.
The Mathematica plot of the $2$-to-$3$ splitting giant graviton interaction is shown in Fig. \ref{fig:evolutionofgiants} for the potential $\phi_{2,3}(z^i)$.     

 Before closing this section, 
 we discuss the necessary and sufficient condition for the existence of instantons 
 when $J \gg 1$. 
 In the case of instantons in the BMN matrix model, 
 the necessary and sufficient condition given in \cite{Bachas:2000dx} 
 has been reproduced by the linearity of the $3$d Laplace equation 
 and the positivity of angular momenta \cite{Kovacs:2015xha}. 
 Since the proof concerning the condition
 does not depend on the dimensionality, 
 we can apply it directly to our case. 
 We conjecture that 
 the condition derived in \cite{Kovacs:2015xha}  
 coincides with the necessary and sufficient condition in our case  
 as well, 
 and we just state the condition: 
 We consider an instanton interpolating $m$ giant gravitons at $t=-\infty$   
 and $n$ giant gravitons at $t=\infty$ 
 characterised by 
 $\mathbb{J}_1 \oplus \cdots \oplus \mathbb{J}_m$ and 
 $\mathbb{J}'_1 \oplus \cdots \oplus \mathbb{J}'_n$, 
 respectively and satisfying $J_1 + \cdots + J_m = J'_1 + \cdots + J'_n = J$ and $m\le n$. 
 Since the angular momenta are positive,  
 one can consider a histogram of $J_i$'s.  
 Drawing a horizontal line at $\tilde J \ge 0$ on the histogram,  
 we define the area of the histogram of $J_i$'s below $\tilde J$: 
 \[
 \mathcal{A}(\tilde J; J_1, \cdots,J_m)\ . 
 \label{eq:areaofhistogram}
 \]   
 As for a histogram of $J'_i$'s, one can define the area $\mathcal{A}(\tilde J; J'_1, \cdots, J'_n)$
 in the same manner. 
The necessary and sufficient condition for the existence of instantons 
can be given \cite{Kovacs:2015xha}:
 \[
\mathcal{A}(\tilde J; J'_1, \cdots, J'_n) \ge \mathcal{A}(\tilde J; J_1, \cdots,J_m), \ \ \ 
\forall \tilde J \ . 
\label{eq:necessaryandsufficientcondition}  
 \]

\section{Giant graviton correlators in CFT}
\label{sec:giantgravitoncorrelatorsincft}
We consider splitting interactions of (concentric) sphere giants  in the dual CFT, {\it i.e.}
$\mathcal{N}=4$ $U(N)$ SYM, in the large-R charge sector \cite{Berenstein:2002jq}. 
The CFT operators dual to giant gravitons with angular momentum $J$ 
are Schur operators of degree $J$ for the unitary group $U(N)$ defined by  \cite{Corley:2001zk,Balasubramanian:2001nh}: 
\[
\chi_{R_J}(Z)={1\over J!}\sum_{\sigma\in S_J}\chi_{R_J}(\sigma)\ 
Z^{i_1}_{i_{\sigma(1)}}Z^{i_2}_{i_{\sigma(2)}}\cdots Z^{i_J}_{i_{\sigma(J)}}
\ ,
\label{eq:dualcftoperator}
\]
where 
$R_{J}$ is an irreducible representation of $U(N)$ expressed by a Young diagram with $J$ boxes, 
$\chi_{R_J} (\sigma)$ is the character of the symmetric group $S_J$ in the representation $R_J$, 
the sum is over  all elements of $S_{J}$ and
$Z$ is an $N$$\times$$N$ complex matrix 
with $i_1, i_2, \cdots , i_J = 1, 2, \cdots , N$.   

If the representation $R_J$ is symmetric (antisymmetric), 
the operator (\ref{eq:dualcftoperator}) corresponds to 
an $AdS$ giant (a sphere giant) \cite{Corley:2001zk,Balasubramanian:2001nh}. 
We will discuss correlation functions of 
Schur operators in antisymmetric representations in order to compare them with the instanton results found in the previous section.\footnote{For more recent progress in the understanding of Schur correlators beyond the $1/2$-BPS sector, see \cite{deMelloKoch:2016whh} and references therein.}

\subsection{Three-point functions of sphere giants}
\label{sec:threegiantcorrelators}
The normalisation of higher point functions can be provided 
by the two point function:
\[
\langle\chi_{A_J}(Z)\chi_{A_J}(\bar{Z})\rangle 
={J!\mbox{Dim}_N(A_J)
\over d_{A_J}}
\ ,
\]
where $A_J$ denotes the antisymmetric representation. 
For antisymmetric representations the dimension $d_{A_{J}}$ of the representation $A_{J}$ is always 1. 
The dimension of the representation $A_J$ of the unitary group is
\[
\mbox{Dim}_N(A_J)={1\over J!}\sum_{\sigma\in S_J}\chi_{A_J}(\sigma)N^{C(\sigma)}\ ,
\]
with $C(\sigma)$ being the number of cycles in the permutation $\sigma$. 
For anti-symmetric representations the character $\chi_A(\sigma)$ is either $1$ or $-1$. 
It is known that 
\[
f_R := {n!\mbox{Dim}_N(R)\over d_R}=\prod_{i,j}(N-i+j)\ ,
\]
where the indices $i$ and $j$ are the label of rows and columns, respectively, 
in the Young diagrams associated with the representation $R$. 
If $R$ is an anti-symmetric representation, there is only one column and $J$ rows, yielding
\[
f_{A_J}=\prod_{i=1}^J(N-i+1)={N!\over (N-J)!}\ .
\]
Thus we have
\[
\langle\chi_{A_J}(Z)\chi_{A_J}(\bar{Z})\rangle={N!\over (N-J)!}\ .
\]
This provides the normalisation of higher point  functions.

Now the 3pt function of sphere giants, corresponding to  
one giant with momentum $J=J_1+J_2$ spliting into two giants with momenta $J_1$ and $J_2$, is given by the formula:
\[
\langle\chi_{A_{J_1}}(Z)\chi_{A_{J_2}}(Z)\chi_{A_J}(\bar{Z})\rangle
=g(A_{J_1}, A_{J_2}; A_{J}) {J!\mbox{Dim}_N(A_J)\over d_{A_{J_1}}d_{A_{J_2}}d_{A_J}}
={N!\over (N-J)!}\ ,
\]
where $g(A_{J_1}, A_{J_2}; A_{J})$ is a Littlewood-Richardson coefficient, an analogue of the Clebsch-Gordan coefficient, and denotes the multiplicity of the representation $A_J$ in the tensor product of representations $A_{J_1}$ and $A_{J_2}$, 
and we have used $g(A_{J_1}, A_{J_2}; A_{J})=1$. 
This is incidentally identical to the two-point function. 
Thus the normalised three-point functions yield
\[
\frac{\langle\chi_{A_{J_1}}(Z)\chi_{A_{J_2}}(Z)\chi_{A_J}(\bar{Z})\rangle}{||\chi_{A_{J_1}}(Z)||\,||\chi_{A_{J_2}}(Z)||\,||\chi_{A_J}(\bar{Z})||}=\sqrt{(N-J_1)!(N-J_2)!\over (N-J)!N!}\ ,
\]
where $||\chi_{A_J}|| := \sqrt{ J! \mbox{Dim}_N(A_J)/d_{A_J} }$. 
In the pp-wave limit, as we discussed in the end of Section \ref{sec:instantonequations}, this exactly agrees with the instanton amplitude as in (\ref{eq:3pointantisymmetricschurs}).

\subsection{General $m$-to-$n$  functions of sphere giants}
\label{sec:generalntomgiantcorrelators}

The general $m\to n$ correlators are also known and given by the formula  \cite{Corley:2001zk,Balasubramanian:2001nh}
\[
&\langle\chi_{R_1}(Z)\cdots \chi_{R_n}(Z)\chi_{T_1}(\bar{Z})\cdots \chi_{T_m}(\bar{Z})\rangle\nn\\
&\hspace{1cm}=\sum_{U}{g(R_1,R_2,\cdots, R_n; U)\over\prod_{i=1}^n d_{R_i}}
{n_U!{\rm Dim}_N(U)\over d_U}{g(T_1,T_2,\cdots, T_m; U)\over\prod_{i=1}^m d_{T_i}}\ .
\]
We only consider the case where all $R$'s and $T$'s are antisymmetric representations. The numbers of boxes for $R_i$ and $T_i$ are $J'_i$ and $J_i$, respectively and $J_1+\cdots+J_m=J'_1+\cdots+J'_n$. 
At large $N$ and $J$ the middle factor  
$
f_U:= {n_{U}!{\rm Dim}_N(U)\over d_{U}}=\prod_{i,j}(N-i+j)
$
is dominated by the representations $U_{\ast}$ which have the largest number of columns as $j$ labels the columns. 
Thus $U_{\ast}$ must have ${\rm min}(n,m)$ columns since it has to be constructible both from $R$'s and $T$'s. 
Without loss of generality we can assume that $m\le n$. 

We first order $R$'s and $T$'s such that the number of boxes $J'_1\ge J'_2\ge\cdots\ge J'_n$ and $J_1\ge J_2\ge\cdots\ge J_m$. 
Then the dominant Young diagrams $U_{\ast}$ at large $N$ and $J$ are composed by first gluing $m$ columns of diagrams $T$'s in this order and then moving some of the boxes down to the left while keeping the number of columns to be $m$. 
The boxes have to be moved so that $U_{\ast}$ is also constructible from $R_1\otimes R_2\otimes\cdots\otimes R_n$.
For these representations we have
\[
f_{U_{\ast}}
:=
{n_{U_{\ast}}\!!\,{\rm Dim}_N(U_{\ast})\over d_{U_{\ast}}}
=\prod_{k=1}^m\prod_{i_k=1}^{J_{k}^{\ast}}(N-i_k+k)=\prod_{k=1}^n{(N+k-1)!\over (N-J_{k\ast}+k-1)!}\ ,
\]
where $J_{k}^{\ast}$ is the number of boxes in the $k$-th column of the Young diagram $U_{\ast}$.
For large $J_k$'s and $N$ we can approximate $J_{k}^{\ast}$'s by $J_k$'s. 
Since the Littlewood-Richardson coefficients are of order 1, their contributions are negligible at large $N$ and $J$,
and we find that
\[
&\frac{\langle\prod_{i=1}^n\chi_{R_i}(Z)\prod_{k=1}^m\chi_{T_k}(\bar{Z})\rangle}
{\prod_{i=1}^n||\chi_{R_i}(Z)||\prod_{k=1}^m||\chi_{T_k}(\bar{Z})||}\nn\\
&\hspace{0.5cm}\simeq\sqrt{(N-J_1)!\cdots (N-J_m)! (N-J'_1)! \cdots  (N-J'_n)!\over (N!)^{n+m}}
\prod_{k=1}^m{(N+k-1)!\over (N-J_k+k-1)!} \notag \\ 
&\hspace{0.5cm} \cong 
e^{{1\over 4N}\left(\sum_{i=1}^n J'^2_i-\sum_{i=1}^{m}J^2_i \right)}
\]
which exactly agrees with the instanton amplitude $e^{-S_E}$ for generic $m$-to-$n$ instanton action (\ref{eq:instantonactionntom}).

\section{The Basu-Harvey equation}
\label{sec:m2m5basuharveyequation}

As we have seen in Section \ref{sec:4dlaplaceequationinriemannspaces}, 
the instanton equation (\ref{eq:instantoneqi}) in the IIB plane-wave matrix model 
can be mapped to the Basu-Harvey equation (\ref{eq:basuharvey}) 
by a change of variables (\ref{eq:changevariables}). 
In order to conform to the original parameterisation in \cite{Basu:2004ed}, 
we make a slight adjustment to the transformation (\ref{eq:changevariables}), 
\[
X^i(t) 
= \sqrt{\frac{\mu \lambda M^2_{11}}{32\pi^3 RT}} \ e^{-\mu t} Z^i(s)\ , \ \ \ 
s=\frac{1}{M_{11}}e^{-2\mu t}\ ,
\label{eq:changevariables2} 
\]
where $M_{11}$ is the eleven-dimensional Planck mass 
and $\lambda$ is the dimensionless coupling constant. 
The instanton equation (\ref{eq:instantoneqi}) 
then becomes\footnote{
The constant matrix $G_5$ introduced in \cite{Basu:2004ed} is slightly different from $\Upsilon_5$, 
but this fact does not spoil the main argument shown in this paper. 
}    
\[
\frac{\text{d} Z^i}{\text{d} s} 
+ \frac{\lambda M^3_{11}}{8 \pi} 
\epsilon^{ijkl} 
\frac{1}{4!}
[\Upsilon_5,Z^j,Z^k,Z^l] =0
\ . 
\label{eq:basuharvey2}
\]
This was proposed as an equation describing M$2$-branes ending on M$5$-branes 
by the M$2$-brane worldvolume theory. 
This is a natural generalisation of Nahm's equation 
describing monopoles or the D$1$-D$3$ system.   
The four scalars $Z^i$'s are $U(J)$ matrices and the coordinates transverse to M$5$-branes,
and $s$ is one of the worldvolume coordinates of M2-branes. 
In the large-$J$ limit a prototypical solution to the Basu-Harvey equation (\ref{eq:basuharvey2}) is a spike made of a bundle of $J$ M$2$-branes 
on a single M$5$-brane of topology, 
$\mathbb{R}_t\times(\mathbb{R}^+_s\times\,{\rm fuzzy}\,\, S^3)\times S_M^1$,  
where $\mathbb{R}_t$ is the time, $\mathbb{R}^+_s$ a semi-infinite line $s\in [0, +\infty]$ and
$S_M^1$ is the M-theory circle corresponding to $\Upsilon_5$. 
In \cite{Howe:1997ue} this was called the ridge solution describing a self-dual string soliton.

When the matrix size $J$ is large, as outlined in (\ref{eq:matrixregularisation1}) - (\ref{eq:matrixregularisation4}), 
the quantum Nambu 4-bracket is replaced by the (classical) Nambu 3-bracket and 
the Basu-Harvey equation becomes
\[
\frac{\partial z^i}{\partial s} 
= 
- \frac{[\sigma]}{3! J_{\lambda}} 
\epsilon^{ijkl} 
\{ 
z^j,z^k,z^l
\} \ ,  
\label{eq:continuumbasuharvey2}
\] 
where 
\[
J_{\lambda} :=  \frac{64 \pi^3 J}{\lambda M^3_{11}}\ . 
\label{eq:jlambda}
\]
By the hodograph transformation (\ref{eq:hodographtr}) 
we solve $s$ as a function of $z^i$'s as done before 
and the equation (\ref{eq:continuumbasuharvey2}) 
can be locally mapped to the 4d Laplace equation. 
Note that the total flux in this case is not $J$ but $J_{\lambda}$ 
(see Appendix \ref{sec:ap:derivationoflaplaceequation} for details).

The aforementioned ridge or spike solution is simply a Coulomb potential in $\mathbb{R}^4$ which is a solution to the 4d Laplace equation:
\[
s= \frac{J_{\lambda}}{4\pi^2 |z^i - a^i|^2}\ , \label{eq:coulomb2}
\] 
with $a^i$ being a constant vector. As remarked, this describes the space $\mathbb{R}^+_s\times S^3$ and the radius of the three-sphere varies along the semi-infinite line as
\[
|z^i - a^i| = \frac{\sqrt{J_{\lambda}}}{2\pi \sqrt{s}}\ . 
\label{eq:ridgesolution}
\]
Note that $s=0$ corresponds to the location of the M$5$-brane 
at which the radius of $S^3$ becomes infinite. 
This is interpreted as an M2-brane spike threading out from a single M5-brane.

We next consider M2-branes stretched between two M5-branes discussed in \cite{Basu:2004ed, Howe:1997ue, Niarchos:2012pn, Sakai:2013gga}.  
The semi-infinite line $\mathbb{R}^+_s$ must be replaced by a finite interval $I_s=\{s|s\in[-s_0, +s_0]\}$
and near the two M$5$-branes at $s=\pm s_0$ the solution behaves as
\[
|z^i-a^i| 
\simeq{\sqrt{J_{\lambda}} \over2 \pi\sqrt{ s\pm s_0}}\ . 
\label{eq:funnelboundarycondition}
\]
An important observation is that the solution with this boundary condition cannot be constructed from Coulomb potentials. 
The reason is that the presence of a point charge necessarily develops a spike as we can see in \eqref{eq:coulomb2}: At the location of the charge $z^i = a^i$, $s$ goes to infinity and thus any solution with point charges cannot represent a finite interval.
This implies that the solutions describing two or more M5-branes are not in the same class of solutions as those describing giant graviton interactions. 
However, similar to the giant graviton case, the idea is to look for solutions to the 4d Laplace equation in the multi-sheeted Riemann space. In this case we expect that the number of sheets corresponds to the number of M5-branes. 

To find the solution which satisfies the boundary condition (\ref{eq:funnelboundarycondition}), 
recall the contour integral expression of the electrostatic potential
\[
\phi(z^i)
={J_{\lambda}\over 16\pi^3}
\oint_{C_{\theta}} \text{d}\theta'{\mathcal{R}^{-2}(e^{i\theta'}\to\zeta'^2)\over 1-e^{i(\theta-\theta')/2}}
{\cosh\rho-\cos\theta\over\cosh\rho-\cos\theta'}\ ,
\label{eq:coulombcontourintegral}
\]
where ${\cal R}^{-2}$ is the 4d Coulomb potential as previously defined in Section \ref{sec:4dlaplaceequationinriemannspaces}.

We can add a constant $c$ to the Coulomb potential
\[
\mathcal{R}^{-2}(e^{i\theta'}\to\zeta'^2)\to \mathcal{R}^{-2}(e^{i\theta'}\to\zeta'^2)+c\ ,
\label{eq:addconstant}
\]
since the constant potential solves the 4d Laplace equation. We now focus on the constant part of the potential
\[
\phi_0(z^i)
={J_{\lambda}\over 16\pi^3}
\oint_{C_{\theta}} \text{d}\theta'{c\over 1-e^{i(\theta-\theta')/2}}
{\cosh\rho-\cos\theta\over\cosh\rho-\cos\theta'}\ .
\label{eq:constantpotential}
\]
Besides the poles at $\theta'=\theta+4k\pi$ with $k\in\mathbb{Z}$, there are poles at 
\[
\theta'=\pm i\rho+2k\pi\ . 
\label{eq:poles}
\]
We deform the contour $C_{\theta}$ to a rectangle of width $4\pi$ (for the two-sheet case) and an infinite height 
while avoiding the poles at $\theta'=\pm i\rho$ and $\pm i\rho+2\pi$. 
Noticing that near the poles
\[
\cosh\rho-\cos\theta'\sim \pm i\sinh\rho\left(\theta'-(\pm i\rho+2k\pi)\right)\ ,
\label{eq:poles2}
\]
similar to the Coulomb potential case, 
the contribution from the first sheet to the constant potential can be found as
\[
\phi^{k=0}_0(z^i)
&=-{cJ_{\lambda}\over 16\pi^3}\oint_{C_{i\rho}+C_{-i\rho}} \text{d}\theta'{1\over 1-e^{i(\theta-\theta')/2}}
{\cosh\rho-\cos\theta\over\cosh\rho-\cos\theta'}\nn\\
&=-{cJ_{\lambda}\over 8\pi^2} {\cosh\rho-\cos\theta\over\sinh\rho}\left({1\over 1-e^{i(\theta-i\rho)/2}}-{1\over 1-e^{i(\theta+i\rho)/2}}\right)\nn\\
&={cJ_{\lambda}\over 8\pi^2} \left[1+{\cos{\theta\over 2}\over \cosh{\rho\over 2}}\right]\ . 
\label{eqnewsoltobasuharvey}
\]
One can check that this solves the 4d Laplace equation. The contribution from the second sheet is $\phi^{k=1}_0(z^i)=c-\phi^{k=0}_0(z^i)$.
Note that at the two asymptotic infinities $(\rho, \theta)\to (0, 0)$ and $(\rho, \theta)\to (0, 2\pi)$ where $z^i$'s go to infinity, 
the electrostatic potential $\phi^{k=0}_0(\vec{z})$ approaches different values, $cJ_{\lambda}/(4\pi^2)$ and 0, respectively. 
By shifting the potential by a constant $s_0$, 
these values can be shifted to $s_0$ and $-s_0$ with the choice $s_0=cJ_{\lambda}/(8\pi^2)$. 
Hence, the potential $\phi^{k=0}_0(\vec{z})$ describes a finite interval of length $2s_0$.

In the $n$-sheeted Riemann space  the trivial constant potential splits into nontrivial potentials defined on each sheet by the contour deformation:
\be
c=\phi^{k=0}_0(z^i)+\phi^{k=1}_0(z^i)+\cdots + \phi^{k=n-1}_0(z^i)\ .
\ee
The explicit form of the potentials for higher $k$'s can be found in the end of this section.

\subsection{M2-branes stretched between two M5-branes -- funnel solution}
\label{sec:m2m5funnel}

\begin{figure}[h!]
\centering \includegraphics[height=2.in]{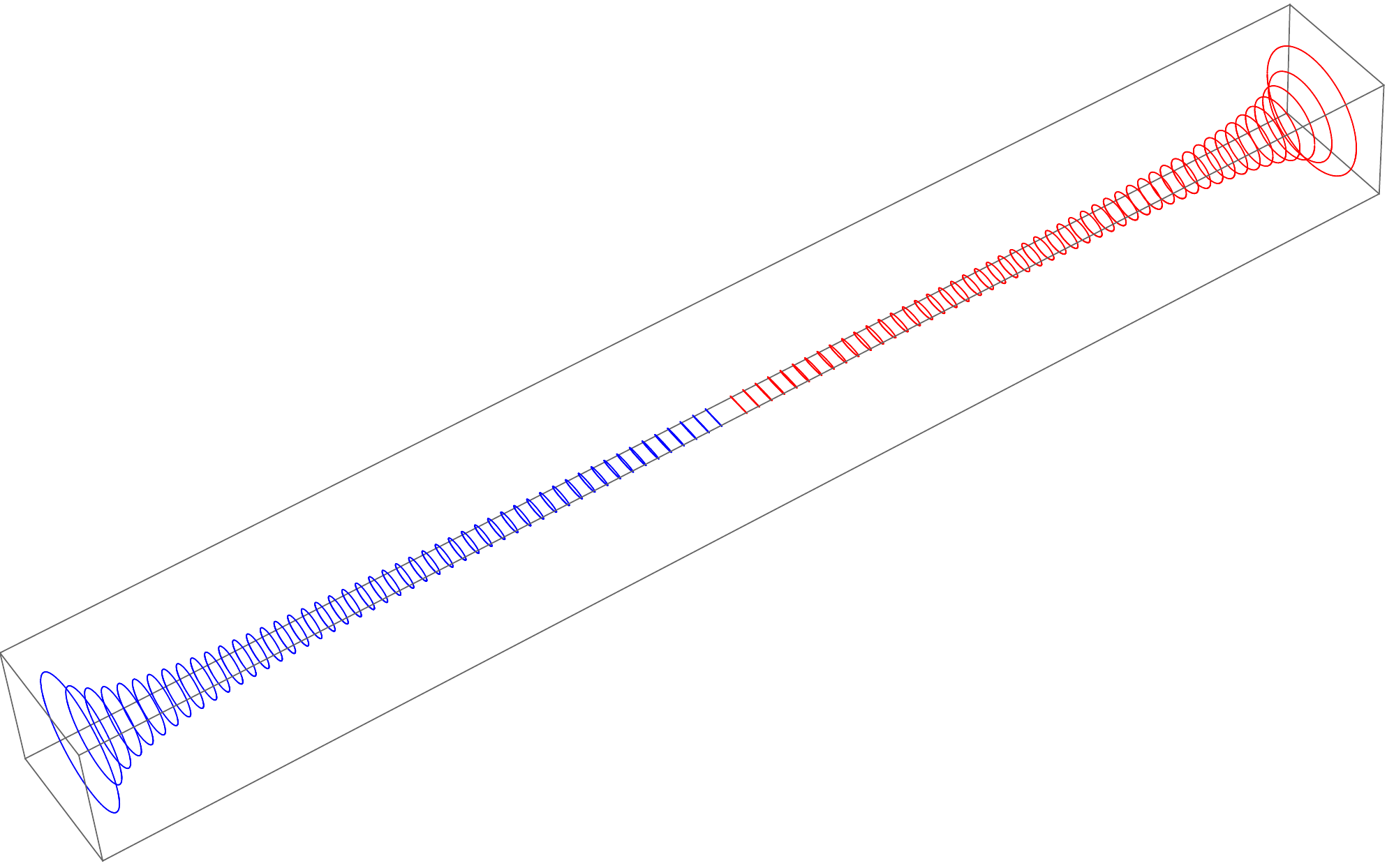} 
\caption{The funnel solution: 
The two ends at $s=\pm s_0$ are the locations of the two M5-branes. 
Each ring is a constant $s$ hypersurface and represents a squashed $S^3$ 
whose radius blows up at the ends and which collapses to a three-ball at the midpoint.}  
\label{funnel}
\end{figure}

As discussed above, the solution representing M$2$-branes stretched between two M$5$-branes can be constructed from a trivial constant electrostatic potential by distilling the contribution from one of the two Riemann sheets.\footnote{The funnel solution has been constructed from different descriptions of the M2-M5 sytem in \cite{Howe:1997ue, Niarchos:2012pn, Sakai:2013gga}.} The M2-branes connecting the two M5-branes have the shape of a funnel: 
\[
s=\phi^{k=0}_0(z^i)-s_0= {s_0\cos{\theta\over 2}\over \cosh{\rho\over 2}}\equiv \phi_{\rm funnel}(z^i)\ . 
\label{eq:sourcelesspotential}
\]
Let us examine this solution more in detail. 
Recalling the parametrisation of the coordinates
\[
\rho=\half\ln{(\xi+a)^2+\eta^2\over (\xi-a)^2+\eta^2}\ ,\qquad
\cos\theta={\xi^2+\eta^2-a^2\over \sqrt{\left((\xi+a)^2+\eta^2\right)\left( (\xi-a)^2+\eta^2\right)}}\ ,
\label{eq:rhoandcostheta}
\]
this can be expressed as
\[
\phi_{\rm funnel}(z^i)=s
=\pm s_0\sqrt{1-{4a^2\over \left(\sqrt{(\xi+a)^2+\eta^2}+\sqrt{(\xi-a)^2+\eta^2}\right)^2}}\ .
\label{eq:sourcelesspotential2}
\]
The midpoint of the funnel $s=0$ corresponds to $\theta=\pi, 3\pi$ which implies $\eta=0$  and $|\xi|\le a$. This is the brach ball $B_3$ and thus in terms of $z_i$'s the midpoint $s=0$ corresponds to a three-ball of radius $a$. 
We plot the funnel solution in Fig. \ref{funnel}. 
The constant $s$ hypersurfaces are squashed three-spheres 
and the radius blows up at the endpoints $s=\pm s_0$ 
and the squashed $S^3$ collapses to a three-ball at $s=0$.\footnote{
If it were in one less dimensions, a squashed $S^2$ would have collapsed or flattened to a $D_2$.}
This collapse of the funnel throat is similar to what happens to D$1$-branes 
stretched between two D$3$-branes \cite{Constable:1999ac}.

Note that at the two asymptotic infinities where $z^i$'s are very large, the coordinates $\xi$ and $\eta$ become very large, since $z_1^2+z_2^2+z_3^2+z_4^2=\xi^2+\eta^2$. Thus the funnel at large $z^i$'s behaves as 
\begin{align}
s\mp s_0\simeq \mp{s_0a^2\over 2|z^i|^2}\qquad\quad (-s_0\le s\le s_0)\ ,
\end{align}
satisfying the boundary condition (\ref{eq:funnelboundarycondition}).

\subsection{M2-branes ending on multiple M5-branes}
\label{sec:m2m5multiplem5s}

The power of this method, 
albeit only in the limit of an infinite number of M$2$-branes, 
is that the solution can be easily generalised to the cases with more than two M$5$-branes. 
We start from the contour integral for a constant potential:
\[
\phi_0(z^i)
={J_{\lambda}\over 8n\pi^3}\oint_{C_{\theta}} \text{d}\theta'{c\over 1-e^{i(\theta-\theta')/n}}
{\cosh\rho-\cos\theta\over\cosh\rho-\cos\theta'}\ . 
\label{eq:coulombmultiplesheets}
\]
Besides the poles at $\theta'=\theta+2nk\pi$ with $k\in\mathbb{Z}$, 
there are poles at 
\[
\theta'=\pm i\rho+2k\ . 
\label{eq:multiplesheetpoles}
\]
We deform the contour $C_{\theta}$ to a rectangle of width $2n\pi$ and an infinite height while avoiding the poles at $\theta'=\pm i\rho+2k\pi$ with $k=0,1,\cdots, n-1$. Noticing that near the poles
\[
\cosh\rho-\cos\theta'\sim \pm i\sinh\rho\left(\theta'-(\pm i\rho+2k\pi)\right)\ ,
\label{eq:multiplesheetpoles2}
\]
similar to the Coulomb potential case, 
the contribution from the first sheet to the constant potential is given by
\[
\phi^{k=0}_0(z^i)
&=-{cJ_{\lambda}\over 8n\pi^3}\oint_{C_{i\rho}+C_{-i\rho}} \text{d}\theta'{1\over 1-e^{i(\theta-\theta')/n}}
{\cosh\rho-\cos\theta\over\cosh\rho-\cos\theta'}\nn\\
&=-{cJ_{\lambda}\over 4n\pi^2} {\cosh\rho-\cos\theta\over\sinh\rho}\left({1\over 1-e^{i(\theta-i\rho)/n}}-{1\over 1-e^{i(\theta+i\rho)/n}}\right)\nn\\
&= {s_0\sinh{\rho\over n}(\cosh\rho-\cos\theta)\over 2n\sinh\rho\left(\cosh^2{\rho\over 2n}-\cos^2{\theta\over 2n}\right)}\ , 
\label{eq:firstsheetmultiplem5}
\]
where $s_0=cJ_{\lambda}/(4\pi^2)$. This asymptotes to $s_0$ at $(\rho, \theta)=(0, 0)$ on the first sheet $k=0$ and $0$ at $(\rho, \theta)=(0, 2k\pi)$ with $k=1,\cdots, n-1$ on the other sheets, corresponding to one M5-brane at $s=s_0$ and $n-1$ M5-branes at $s=0$. 

The general solutions are given by the superposition of the potentials from different sheets. For example, the superposition of the two  $\phi^{k=0}_0(z^i)$ and $\phi^{k=1}_0(z^i)$
\[
\phi_0(z^i) 
= {s_1\sinh{\rho\over n}(\cosh\rho-\cos\theta)\over 2n\sinh\rho\left(\cosh^2{\rho\over 2n}-\cos^2{\theta\over 2n}\right)}
+{s_2\sinh{\rho\over n}(\cosh\rho-\cos\theta)\over 2n\sinh\rho\left(\cosh^2{\rho\over 2n}-\cos^2{(\theta+2\pi)\over 2n}\right)}
\label{eq:newsolutobasuharveymultiplem5}
\]
asymptotes to $s_1$ at $(\rho, \theta)=(0, 0)$ on the first sheet, $s_2$ at $(\rho, \theta)=(0, 2(n-1)\pi)$ on the $n$-th sheet and $0$ on the other sheets, corresponding to one M5-brane at $s=s_1$, another M5-brane at $s=s_2$ and $n-2$ M5-branes at $s=0$.  

\begin{figure}[h]
\centering
\includegraphics[width=4.3in]{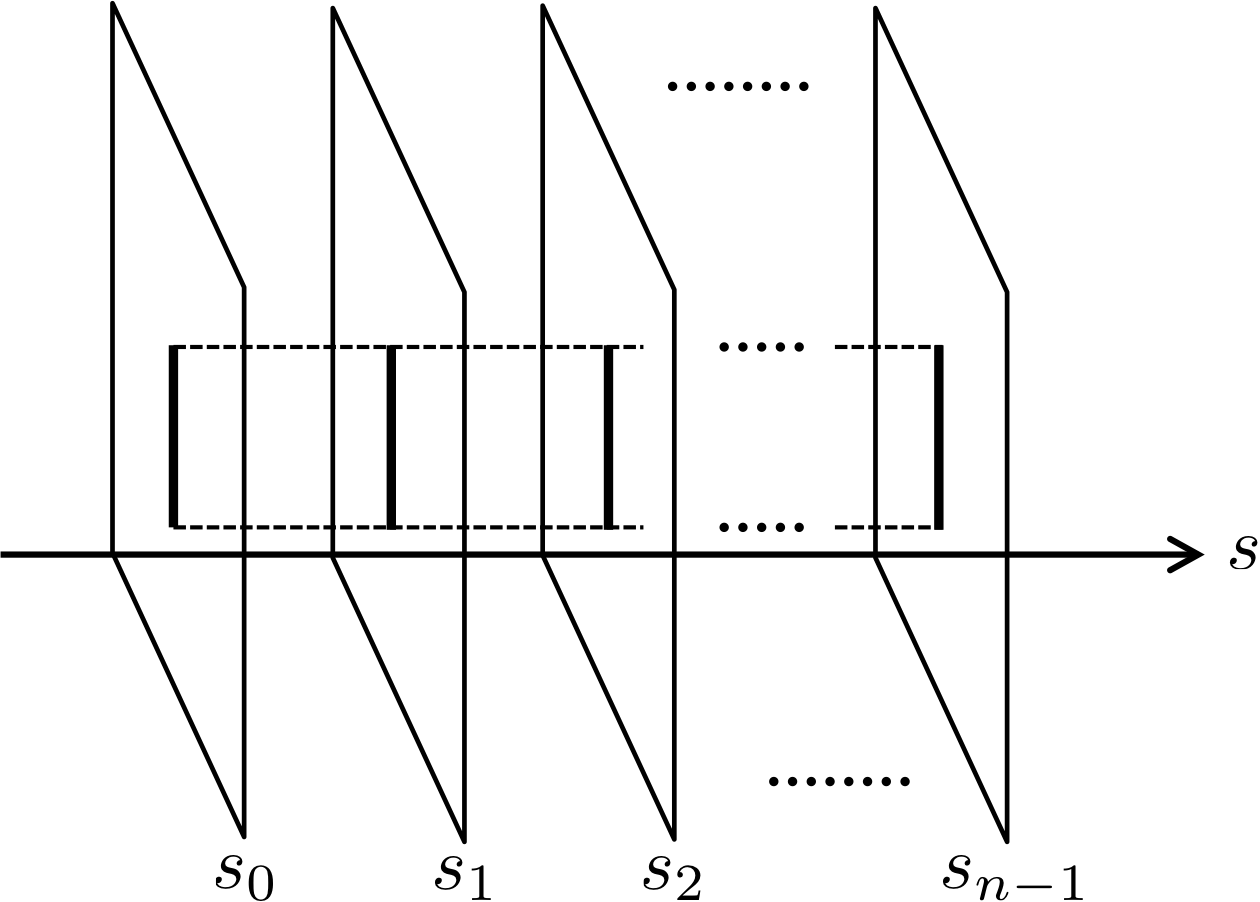}
\caption{M$5$-branes are located at $s=s_k$ with $k=0,1,\cdots,n-1$ labelling the sheets of the Riemann space. The thick line segments represent the branch three-balls and are all identified. M2-branes ending on multiple M5-branes correspond to the electrostatic potential distilled from a constant potential by means of contour deformation and there are no charges present in the Riemann space. M2-branes connecting M5-branes all meet at the branch three-balls.}
\label{fig:m5s}
\end{figure}

\begin{figure}[h!]
\centering \includegraphics[height=2.in]{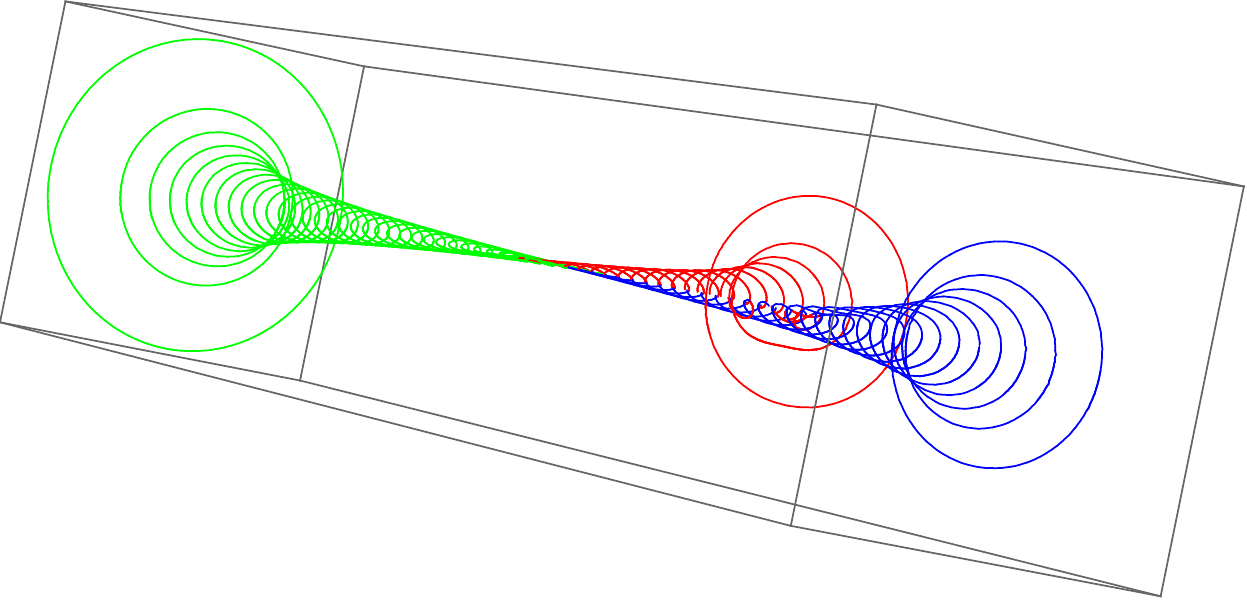} 
\caption{The M$2$-branes junction ending on three different M$5$-branes.}  
\label{funnel3}
\end{figure}

We can construct the most general solution with all different asymptotic values describing $n$ separated M5-branes:
\[
\phi_0(z^i)=\sum_{k=0}^{n-1}{s_k\sinh{\rho\over n}(\cosh\rho-\cos\theta)\over 2n\sinh\rho\left(\cosh^2{\rho\over 2n}-\cos^2{(\theta+2(n-k)\pi)\over 2n}\right)}\ ,
\label{eq:newsolutobasuharveymultiplem5generic}
\]
where $s_k$ is the modulus representing the location of each M5-brane (see Fig.\ref{fig:m5s}). 
As an example of the cases with more than two M$5$-branes, 
we plot an M$2$-branes junction ending on three different M$5$-branes 
corresponding to $n=3$ with some choice of the locations $(s_1,s_2,s_3)$ in Fig. \ref{funnel3}.

\section{Summary and discussions} 
\label{sec:discussion}

We studied the dynamical process of giant gravitons, 
\textit{i.e.} their splitting and joining interactions, 
in the type IIB string theory on $AdS_5 \times S^5$. It was made possible by restricting ourselves to  
small size giants whose angular momenta are in the range $N^{1/2}\ll J\ll N$ for which the spacetime can be well approximated by the plane-wave background. 
We found that the most effective description was provided by the tiny graviton matrix model of Sheikh-Jabbari \cite{SheikhJabbari:2004ik, SheikhJabbari:2005mf}, which we referred to as the IIB plane-wave matrix model, rather than BMN's type IIB string theory on the pp-wave background.

We showed, in particular, that their splitting/joining interactions can be described by 
instantons/anti-instantons in the IIB plane-wave matrix model. 
They connect one vacuum, a cluster of $m$ concentric (fuzzy) sphere giants, in the infinite past to another vacuum, a cluster of $n$ concentric (fuzzy) sphere giants, in the infinite future. 
In the large $J$ limit the instanton equation can be mapped locally to the 4d Laplace equation
and the $m$-to-$n$ interaction corresponds to the Coulomb potential of $m$ point charges on an $n$-sheeted Riemann space.

Giant graviton interactions are dual to correlators of Schur polynomial operators in ${\cal N}=4$ SYM. The latter have been calculated exactly by Corley, Jevicki and Ramgoolam \cite{Corley:2001zk}. We compared the instanton amplitudes to the CFT correlators and found an exact agreement for generic $m$ and $n$ within the validity of our approximation. This lends strong support for our description of giant graviton interactions. 
However, to be more precise, the agreements are only for the sphere giants which expand in $S^5$ and are dual to antisymmetric Schur operators and a puzzle, as pointed out in \cite{Takayanagi:2002nv}, remains for the AdS giants which expand in $AdS_5$ and are dual to symmetric Schur operators. The issue is that the correlators of symmetric Schur operators exponentially grow rather than damp in the pp-wave limit.

A next step would be going beyond the classical approximation and include fluctuations about (anti-)instantons in order to find $N/J^2$ corrections. This involves integrations over bosonic and fermionic zero modes and requires finding the moduli space of (anti-)instantons which includes geometric moduli associated with the Riemann space, 
\textit{i.e.} the number of sheets and the number, positions and shapes of branch three-balls, as discussed in the case of membrane interactions \cite{Kovacs:2015xha}. This is not an easy problem.

As a byproduct of this study we also found new results on multiple M5-branes. 
We exploited the fact that the instanton equation is identical to the Basu-Harvey equation which describes the system of M$2$-branes ending on M$5$-branes \cite{Basu:2004ed}.
In the large $J$ limit which corresponds, in the Basu-Harvey context, to a large number of M2-branes, we found the solutions describing M2-branes ending on multiple M5-branes, including the funnel solution and an M2-branes junction connecting three M5-branes as simplest examples. The number $n$ of M5-branes corresponds to the number of sheets in the Riemann space, and  
somewhat surprisingly, multiple M5-branes solutions are constructed from a trivial constant electrostatic potential.
Upon further generalisations, for example, adding more branch balls, the effective theory on the moduli space of our solutions might shed light on the low energy effective theory of multiple M5-branes \cite{Ho:2008nn, Lambert:2010wm, Chu:2009iv, Saemann:2017zpd}.

Finally, our technique is applicable to the well-known SU($\infty$) limit of Nahm's equation which describes (an infinite number of) D$1$-branes ending on D$3$-branes by mapping it locally to the 3d Laplace equation \cite{Ward:1989nz,Hoppe:1994np}.
This might give us a new perspective on the moduli space of monopoles.

\section*{Acknowledgement}

We would like to thank 
Robert de Mello Koch, 
Chong-Sun Chu, 
Masashi Hamanaka, 
Satoshi Iso, 
Hiroshi Isono, 
Stefano Kovacs, 
Niels Obers, 
Shahin Sheikh-Jabbari, 
Hidehiko Shimada
and 
Seiji Terashima  
for discussions and 
comments. 
SH would like to thank the Graduate School of Mathematics at Nagoya University, Yukawa Institute for Theoretical Physics and Chulalongkorn University for their kind hospitality. 
YS would like to thank all members of the String Theory Group at the University of the Witwatersrand for their kind hospitality, where this work was initiated. 
The work of SH was supported in part by the National Research Foundation of South Africa and DST-NRF Centre of Excellence in Mathematical and Statistical Sciences (CoE-MaSS).
The work of YS was funded under CUniverse research promotion project by Chulalongkorn 
University (grant reference CUAASC).

\appendix

\section{A derivation of the Laplace equation}
\label{sec:ap:derivationoflaplaceequation}
We are going to show that the following differential equation can be mapped to the $n$-dimensional Laplace equation:
\[
\frac{\partial z^{p}}{\partial s} 
= - \frac{[\sigma]}{(n-1)! J} \epsilon^{p p_1 \cdots p_{n-1}} 
\{ z^{p_1}, z^{p_2}, \cdots, z^{p_{n-1}} \}\ , 
\label{eq:ap:differentialeq}
\]
where $z^p$ and $z^{p_i}$ with $p,\, p_i =1,2,\cdots,n$ are functions of $(s, \sigma^l)$ with $l=1,2,\cdots,n-1$. 
On the RHS the Nambu $(n-1)$-bracket is defined by 
\[
\{ z^{p_1}, z^{p_2}, \cdots, z^{p_{n-1}}  \} 
= \epsilon^{l_1 l_2 \cdots l_{n-1}} 
\frac{\partial z^{p_1}}{\partial \sigma^{l_1}}
\frac{\partial z^{p_2}}{\partial \sigma^{l_2}} 
\cdots 
\frac{\partial z^{p_{n-1}}}{\partial \sigma^{l_{n-1}}}\ , 
\label{eq:ap:nambunminusone}
\]
and 
\[
[\sigma] = \int \text{d}^{n-1}\sigma\ . 
\label{eq:ap:sigmavolume}
\]
The equation (\ref{eq:ap:differentialeq}) describes
an evolution of an $(n-1)$-dimensional hypersurface embedded in $\mathbb{R}^n$ 
with time $s$. 
We can express this hypersurface at a constant time slice as 
a function $\phi(z^1,z^2,\cdots,z^n)$ satisfying the equation
\[
s = \phi (z^1,z^2,\cdots,z^n)\ . 
\label{eq:ap:sisphi}
\]
We now follow the proof in \cite{Kovacs:2015xha} given in the case of $n=3$, extend it to general $n$ and show that the electrostatic potential $\phi(z^1,\cdots, z^n)$ satisfies the $n$-dimensional Laplace equation. 
First note that the $n$-dimensional volume element can be expressed as
\[
\text{d}z^1\wedge \text{d}z^2\wedge \cdots \wedge \text{d}z^n 
&= \epsilon^{p p_1 \cdots p_{n-1}} 
\frac{\partial z^{p}}{\partial s}
\frac{\partial z^{p_1}}{\partial \sigma^1} 
\cdots 
\frac{\partial z^{p_{n-1}}}{\partial \sigma^{n-1}} 
\text{d}s\wedge \text{d}\sigma^1 \wedge \cdots \wedge \text{d}\sigma^{n-1} \notag \\
&=: \frac{\partial z^p}{\partial s} \text{d}s \wedge \text{d}\Sigma^p\ , 
\label{eq:ap:ndvolumeelement} 
\] 
where $p=1,2,\cdots,n$
and from \eqref{eq:ap:sisphi}
\[
1 = \frac{\partial \phi}{\partial z^p} \frac{\partial z^p}{\partial s}\ .
\label{eq:ap:chainrule}
\]
By multiplying \eqref{eq:ap:differentialeq} by ${\partial\phi\over \partial z^p}{\rm d}\sigma^1 \wedge \text{d}\sigma^2 \wedge \cdots \wedge \text{d}\sigma^{n-1}$, 
the equation (\ref{eq:ap:differentialeq}) can then be rewritten as
\[
\frac{J}{[\sigma]} 
\text{d}\sigma^1 \wedge \text{d}\sigma^2 \wedge \cdots \wedge \text{d}\sigma^{n-1}=- \frac{\partial \phi}{\partial z^p}  \ \text{d}\Sigma^p\ . 
\label{eq:ap:prelaplace}
\]
Integrating (\ref{eq:ap:prelaplace}) over the boundary hypersurface $\del V_n=\prod_{i=1}^{n-1} I_i\times \del I_s$
of the infinitesimal volume $V_n=\prod_{i=1}^{n-1} I_i\times I_s$ where the intervals $I_i=[\sigma_i, \sigma_i+d\sigma_i]$ and $I_s=[s, s+ds]$, the flux conservation yields 
\[
0=\int_{\del V_n}  \frac{\partial \phi}{\partial z^p} \ \text{d}\Sigma^p =\int_{V_n}\Delta\phi\, \text{d}z^1\wedge \text{d}z^2\wedge \cdots \wedge \text{d}z^n\ .
\label{eq:ap:laplaceequation}
\]
This is nothing but the $n$-dimensional Laplace equation. 

In order to find $z^p(s,\sigma^l)$ from solutions to the Laplace equation $\Delta\phi(z^1,\cdots,z^n)=0$,  
we use (\ref{eq:ap:prelaplace}) and \eqref{eq:ap:chainrule}. Namely, the equation (\ref{eq:ap:prelaplace}) implies that the electric flux density in the $(n-1)$-dimensional $\sigma$-space is the constant $\frac{J}{[\sigma]}$ at a given $s$. In other words, the Guassian surface of constant electric fields is tangent to the $\sigma$-space and normal to the time $s$: $\vec{E}\cdot{\del\vec{z}\over\del\vec{\sigma}}=0$  and $\vec{E}\cdot{\del\vec{z}\over\del s}=-1$ using  \eqref{eq:ap:chainrule},
where the electric field $\vec{E}(z^1,\cdots,z^n)=-\frac{\partial \phi}{\partial \vec{z}}$. These $n$ equations determine $z^p$'s as functions of $(s, \sigma^l)$.


\section{The Euclidean 3-brane theory}
\label{sec:ap:euclideanthreebranetheory}
In this appendix we are going to show that the continuum version of the Basu-Harvey equation (\ref{eq:continuumbasuharvey}) 
can be obtained from the Euclidean $3$-brane theory. 

We start with the gauge-fixed lightcone Hamiltonian (\ref{eq:lightconehamiltonian}). 
Using Hamilton's equation,   
\[
\frac{\partial x^I}{\partial t} = \frac{[\sigma]}{(-P_-)}p^I = \frac{[\sigma]R}{J} p^I\ , 
\]
the action becomes
\[
I
&= \frac{J}{2R [\sigma]} 
\int \text{dt}\text{d}^3\sigma\
\biggl[
\left( \frac{\partial x^I}{\partial t} \right)^2 
-\mu^2 (x^I)^2 
- \frac{1}{3!} \left( \frac{R T [\sigma] }{J} \right)^2 
\{x^I,x^J,x^K \}^2 \notag \\
& \ \ \ 
+ \frac{\mu R T [\sigma]}{3J} 
\left( 
\epsilon^{ijkl} x^i \{ x^j,x^k,x^l \} 
+ \epsilon^{abcd} x^a \{ x^b,x^c,x^d \} 
\right)
\biggl]\ . 
\label{eq:ap:bosonicthreebraneaction}
\]
By a Wick-rotation the Euclidean action yields
\[
I_E
&= \frac{J}{2R [\sigma]} 
\int \text{dt}\text{d}^3\sigma\
\biggl[
\left( \frac{\partial x^I}{\partial t} \right)^2 
+ \mu^2 (x^I)^2 
+ \frac{1}{3!} \left( \frac{R T [\sigma]}{J} \right)^2 
\{x^I,x^J,x^K \}^2 \notag \\
& \ \ \ 
- \frac{\mu R T [\sigma]}{3J} 
\left( 
\epsilon^{ijkl} x^i \{ x^j,x^k,x^l \} 
+ \epsilon^{abcd} x^a \{ x^b,x^c,x^d \} 
\right)
\biggl]\ , 
\label{eq:ap:euclideanbosonicthreebraneaction}
\]
where $t$ is the Euclidean time. 
The Euclidean action (\ref{eq:ap:euclideanbosonicthreebraneaction}) can be recast as a sum of squares and boundary terms: 
\[
I_E
=
\frac{J}{2R [\sigma]} 
\int \text{d}t\text{d}^3\sigma \ 
&\biggl[
\left( 
\frac{\partial x^i}{\partial t}  
\pm \mu x^i 
\mp 
\frac{RT[\sigma]}{3! J}
\epsilon^{ijkl}\{x^j,x^k,x^l \}
\right)^2 \notag \\
&+\left( 
\frac{\partial x^a}{\partial t}  
\pm \mu x^a 
\mp 
\frac{RT[\sigma]}{3! J}
\epsilon^{abcd}\{x^b,x^c,x^d \}
\right)^2 \notag \\
&+ 
\frac{1}{2}
\left( \frac{RT [\sigma]}{J} \right)^2
\biggl(
  \{x^i,x^a,x^b \}^2 
+  \{ x^a ,x^i,x^j \}^2
\biggl) \notag \\
&\mp \frac{\text{d}}{\text{d}t} 
\left( 
\mu \left( x^i \right)^2 
-  
\frac{JRT[\sigma]}{12}
\epsilon^{ijkl}x^i \{ x^j,x^k,x^l \}
\right) \notag \\
&\mp \frac{\text{d}}{\text{d}t} 
\left( 
\mu \left( x^a \right)^2 
- 
\frac{JRT[\sigma]}{12}
\epsilon^{abcd}x^a \{ x^b,x^c,x^d \}
\right)
\biggl]\ . 
\label{eq:ap:euclideanbosonicthreebraneaction2}
\]
This is minimised when the first order BPS equations are satisfied\footnote{
One can show the non-negativity of the Euclidean action by constructing 
the equations analogous to (\ref{eq:gradientflow}) in the IIB plane-wave matrix model.
}
\[
& \frac{\partial x^i}{\partial t}  
\pm \mu x^i 
\mp 
\frac{RT[\sigma]}{3! J}
\epsilon^{ijkl}\{x^j,x^k,x^l \} =0\ , \ \ \ x^a=0\ , \ \ \ x^i \ne 0\ , \label{eq:continuuminstantoneq} \\
&\frac{\partial x^a}{\partial t}  
\pm \mu x^a 
\mp 
\frac{RT[\sigma]}{3! J}
\epsilon^{abcd}\{x^b,x^c,x^d \}=0\ , \ \ \ x^i=0\ , \ \ \ x^a \ne 0\ , \label{eq:continuuminstantoneq2} \\
&x^i = x^a=0\ . 
\]
By a change of variables,
\[
x^I (t,\sigma^{\mu}) 
= \sqrt{\frac{2\mu}{RT}} e^{-\mu t} z^I (s,\sigma^{\mu})\ ,  
\ \ \ s= e^{-2\mu t}\ ,
\]
the BPS equations (\ref{eq:continuuminstantoneq}) and (\ref{eq:continuuminstantoneq2}) transform to
\[
\frac{\partial z^i}{\partial s} 
= 
\mp \frac{[\sigma]}{3! J} 
\epsilon^{ijkl} 
\{ 
z^j,z^k,z^l
\}
= \mp \frac{[\sigma]}{J}\epsilon^{ijkl} \frac{\partial z^j}{\partial \sigma^1}\frac{\partial z^k}{\partial \sigma^2}\frac{\partial z^l}{\partial \sigma^3}\ ,
\label{eq:contnahm}
\]
\[
\frac{\partial z^a}{\partial s} 
= 
\mp \frac{[\sigma]}{3! J} 
\epsilon^{abcd} 
\{ 
z^b,z^c,z^d
\}
= \mp \frac{[\sigma]}{J}\epsilon^{abcd} \frac{\partial z^b}{\partial \sigma^1}\frac{\partial z^c}{\partial \sigma^2}\frac{\partial z^d}{\partial \sigma^3}\ .
\label{eq:contnahm2}
\]
These equations are the continuum version of the Basu-Harvey equation (\ref{eq:continuumbasuharvey})
and by a hodograph transformation 
they can be locally mapped to the $4$d Laplace equation 
as explained in Appendix \ref{sec:ap:derivationoflaplaceequation}.

\section{Three-spheres and their quantisation}
\label{sec:ap:threespheres}
We give a brief review of the relation between three-spheres and the Nambu $3$-bracket. 
Upon quantisation of this relation, $S^3$'s become 
fuzzy $S^3$'s and the Nambu 3-bracket is replaced by the quantum Nambu  $4$-bracket. 
The construction of fuzzy $S^3$'s will be given below. 
The parameter $\ell$ in the quantisation of the Nambu bracket  is analogous to $\hbar$ in quantum mechanics (\ref{eq:hbariniib}) 
and fixed by the requirement that the radius of $S^3$ coincides with that of fuzzy $S^3$.     

We start with an $S^3$ of radius $r$ 
\[
\sum^{4}_{i=1} (x^i)^2 = r^2\ .  
\label{eq:ap:definingequationofthreesphere}
\] 
We choose the spherical coordinates to be
\[
x^i = r n^i 
= r (\cos \lambda, \sin \lambda \cos \varphi, \sin \lambda \sin \varphi \cos  \omega, \sin \lambda \sin \varphi \sin \omega)\ . 
\label{eq:ap:xbysphericalcoordinates} 
\]
We can then show that $x^i$'s satisfy the following equation:
\[
x^i = \frac{1}{3! r^2} \epsilon^{ijkl} 
\{ x^j,x^k,x^l \}\ , \ \ \ 
\text{with} \ \ \ 
\{ x^i,x^j,x^k \} := 
\epsilon^{lmn} 
\frac{\partial x^i}{\partial \sigma^l} 
\frac{\partial x^j}{\partial \sigma^m}
\frac{\partial x^k}{\partial \sigma^n}\ , 
\label{eq:3sphere}
\]
where $\sigma^l$ ($l=1,2,3$)  
are the coordinates on the $S^3$ and have the volume element
\[
\text{d}^3 \sigma 
= \sin^2 \lambda \sin \varphi \ \text{d}\lambda \text{d}\varphi \text{d} \omega\ . 
\label{eq:qp:so4symmetricvolume}
\]
Here $\{ \ast , \ast, \ast \}$ is the Nambu $3$-bracket. 
For a unit $S^3$, in particular, we have
\[
n^{i} 
= \frac{1}{3!} \epsilon^{ijkl} \{n^j,n^k,n^l \}
= \frac{1}{\sin^2 \lambda \sin \varphi} 
\epsilon^{ijkl} \frac{\partial n^j}{\partial \lambda} \frac{\partial n^{k}}{\partial \varphi} \frac{\partial n^l}{\partial \omega}\ .
\]
This establishes the relation between $S^3$'s and the Nambu 3-bracket.

\subsection{Fuzzy three-spheres} 
\label{subsec:ap:fuzzythreespheres}
The fuzzy $S^3$'s can be constructed 
as a subspace of fuzzy $S^4$'s \cite{Guralnik:2000pb, Ramgoolam:2001zx}. 
We only recapitulate the essential part of the construction and leave details to the original papers \cite{Guralnik:2000pb, Ramgoolam:2001zx}.    
 
We introduce $J \times J$ matrices, 
\[
\Upsilon^i 
&= P_{\mathcal{R}} (\Gamma^i \otimes \mathbf{1}^{n-1})_{\text{sym}} P_{\mathcal{R}}\ , \label{eq:ap:upsiloni} \\
\Upsilon_5 
&= P_{\mathcal{R}} (\Gamma_5 \otimes \mathbf{1}^{n-1})_{\text{sym}} P_{\mathcal{R}}\ , \label{eq:ap:upsilon5}
\]
where $\Gamma^i$ are the four-dimensional $4\times 4$ Dirac matrices, 
$\Gamma_5$ is the $SO(4)$ chirality operator, 
$\mathbf{1}$ is the $4 \times 4$ unit matrix, 
$n$ is an odd integer  
and the suffix `sym' denotes a symmetric $n$-fold tensor product. 
Here $P_{\mathcal{R}}$ is the projector onto the $J \times J$ representation $\mathcal{R}$ 
of $SO(4) \cong SU(2)_L \times SU(2)_R$  
given by 
\[
\mathcal{R} 
= \left(
\frac{n-1}{4}, \frac{n+1}{4}
\right)
\oplus 
\left(
\frac{n+1}{4}, 
\frac{n-1}{4}
\right)\ , 
\label{eq:ap:r}
\] 
where $(j_L,j_R)$ is an irreducible representation of 
$Spin (4) = SU(2)_L \times SU(2)_R$. 
The dimension of $\mathcal{R}$ specifies the size of matrices $J$:
\[
J = \text{dim}\ \mathcal{R} = \frac{(n+1)(n+3)}{2}\ . 
\label{eq:ap:j}
\]
Using $\Upsilon^i$ and $\Upsilon_5$, one can construct 
a fuzzy $S^3$ of unit radius: 
\[
N^i = - \frac{J}{3!} \epsilon^{ijkl} [N^j,N^k,N^l,\Upsilon_5]\ , 
\ \ \ \sum^{4}_{i=1} (N^i)^2 = \mathbf{1}_{J \times J}\ , 
\label{eq:ap:unitfuzzythreesphere}
\]
where the quantum Nambu $4$-bracket is defined in (\ref{eq:quantumnambufourbracket}) 
and   
\[
N^i = \frac{1}{\sqrt{J}} \Upsilon^i \Upsilon_5\ . 
\label{eq:ap:matni}
\]
This can be easily generalised to a fuzzy $S^3$ of radius $r_F$ by
\[
X^i = r_F N^i = \frac{r_F}{\sqrt{J}} \Upsilon^i \Upsilon_5\ ,
\label{eq:ap:matxi}
\]
which satisfy 
\[
X^i = - \frac{J}{3! r^2_F} \epsilon^{ijkl} [X^j,X^k,X^l,\Upsilon_5]\ , 
\ \ \ \sum^{4}_{i=1} (X^i)^2 = r^2_F \mathbf{1}_{J \times J}\ . 
\label{eq:ap:fuzzythreespherewithrf}
\]
We denote the irreducible $J \times J$ representation  (\ref{eq:ap:r}) of $Spin(4)$ by $\mathbb{J}$. 
In the case of a reducible representation, 
\[
\mathbb{J}_1 \oplus \mathbb{J}_2 \oplus \cdots \oplus \mathbb{J}_n
\label{eq:ap:reduciblerepofso4}
\] 
with $J_1 + J_2 + \cdots + J_n = J$, 
the solutions to the equation (\ref{eq:ap:fuzzythreespherewithrf}) 
form $n$ concentric fuzzy $S^3$'s.   
This establishes the relation between fuzzy $S^3$'s and the Nambu 4-bracket.

\subsection{Fixing the quantisation parameter}
\label{sec:ap:quantisngnumbuthreebracket}
We elaborate on our choice of the quantisation parameter $l$ in (\ref{eq:matrixregularisation3}).  
Recall that the three-brane theory defined by the Hamiltonian (\ref{eq:lightconehamiltonian}) 
has the vacua obeying
\[
x^i 
= \frac{1}{3! r^2} \epsilon^{ijkl} \{ x^j,x^k,x^l \}\ , 
\ \ \ x^a=0\ , 
\label{eq:ap:threespherexiex}
\]
where
\[
r= \sqrt{ \frac{ \mu J  }{[\sigma] R T} }\ . 
\label{eq:ap:radiusofthreesphereex}
\]
The solution to (\ref{eq:ap:threespherexiex}) is given by (\ref{eq:ap:xbysphericalcoordinates}) 
which forms an $S^3$ of radius $r$. 
Since the $\sigma$-coordinates are chosen as in (\ref{eq:qp:so4symmetricvolume}), 
we have
\[
[\sigma] = \int \text{d}^3 \sigma
= \int^{\pi}_{0} \text{d}\lambda \int^{\pi}_{0} \text{d}\varphi \int^{2\pi}_{0} \text{d}\omega\ 
\sin^2 \lambda \sin \varphi 
= 2 \pi^2\ . 
\label{eq:ap:volumesigma2pisquare}
\]
As a result the radius (\ref{eq:ap:radiusofthreesphereex}) of the $S^3$ is  found as
\[
r = \sqrt{\frac{\mu J }{2 \pi^2 RT}}
=R_S \sqrt{\frac{J}{N}}\ , 
\label{eq:ap:radiusofthreesphereex2}
\]
where we used (\ref{eq:mutor}), 
(\ref{eq:relationamongrandrs}) 
and (\ref{eq:adscftdictionary}). 
Indeed,  with the choice of $\ell$ in (\ref{eq:hbariniib}),
the quantisation procedure (\ref{eq:matrixregularisation1}) -- (\ref{eq:matrixregularisation4}) yields  
the radius of the fuzzy $S^3$ to be (\ref{eq:radiusofgiant}) 
which coincides with (\ref{eq:ap:radiusofthreesphereex2}).







\begin{thebibliography}{40}


\bibitem{McGreevy:2000cw}
  J.~McGreevy, L.~Susskind and N.~Toumbas,
  ``Invasion of the giant gravitons from Anti-de Sitter space,''
  JHEP {\bf 0006} (2000) 008
  doi:10.1088/1126-6708/2000/06/008
  [hep-th/0003075].

\bibitem{Grisaru:2000zn}
  M.~T.~Grisaru, R.~C.~Myers and O.~Tafjord,
  ``SUSY and goliath,''
  JHEP {\bf 0008} (2000) 040
  doi:10.1088/1126-6708/2000/08/040
  [hep-th/0008015].
  
\bibitem{Hashimoto:2000zp}
  A.~Hashimoto, S.~Hirano and N.~Itzhaki,
  ``Large branes in AdS and their field theory dual,''
  JHEP {\bf 0008} (2000) 051
  doi:10.1088/1126-6708/2000/08/051
  [hep-th/0008016].

\bibitem{Balasubramanian:2001nh}
  V.~Balasubramanian, M.~Berkooz, A.~Naqvi and M.~J.~Strassler,
  ``Giant gravitons in conformal field theory,''
  JHEP {\bf 0204} (2002) 034
  doi:10.1088/1126-6708/2002/04/034
  [hep-th/0107119].  
  

\bibitem{Corley:2001zk}
  S.~Corley, A.~Jevicki and S.~Ramgoolam,
  ``Exact correlators of giant gravitons from dual N=4 SYM theory,''
  Adv.\ Theor.\ Math.\ Phys.\  {\bf 5} (2002) 809
  doi:10.4310/ATMP.2001.v5.n4.a6
  [hep-th/0111222].  


\bibitem{Maldacena:1997re}
  J.~M.~Maldacena,
  ``The Large N limit of superconformal field theories and supergravity,''
  Int.\ J.\ Theor.\ Phys.\  {\bf 38} (1999) 1113
   [Adv.\ Theor.\ Math.\ Phys.\  {\bf 2} (1998) 231]
  doi:10.1023/A:1026654312961, 10.4310/ATMP.1998.v2.n2.a1
  [hep-th/9711200]. 
  

\bibitem{Berenstein:2002jq}
  D.~E.~Berenstein, J.~M.~Maldacena and H.~S.~Nastase,
  ``Strings in flat space and pp waves from N=4 superYang-Mills,''
  JHEP {\bf 0204} (2002) 013
  doi:10.1088/1126-6708/2002/04/013
  [hep-th/0202021].


\bibitem{Gubser:2002tv}
  S.~S.~Gubser, I.~R.~Klebanov and A.~M.~Polyakov,
  ``A Semiclassical limit of the gauge / string correspondence,''
  Nucl.\ Phys.\ B {\bf 636} (2002) 99
  doi:10.1016/S0550-3213(02)00373-5
  [hep-th/0204051].


\bibitem{Frolov:2002av}
  S.~Frolov and A.~A.~Tseytlin,
  ``Semiclassical quantization of rotating superstring in AdS(5) x S**5,''
  JHEP {\bf 0206} (2002) 007
  doi:10.1088/1126-6708/2002/06/007
  [hep-th/0204226].  
   
\bibitem{Penrose1976}
R.~Penrose, ``Any spacetime has a plane wave as a limit,'' 
in \textit{Differential geometry and relativity}, 
pp. 271-275, M.~Cahen and M.~Flato editors (1976). 


\bibitem{Gueven:2000ru}
  R.~Gueven,
  ``Plane wave limits and T duality,''
  Phys.\ Lett.\ B {\bf 482} (2000) 255
  doi:10.1016/S0370-2693(00)00517-7
  [hep-th/0005061].


\bibitem{Blau:2002dy}
  M.~Blau, J.~M.~Figueroa-O'Farrill, C.~Hull and G.~Papadopoulos,
  ``Penrose limits and maximal supersymmetry,''
  Class.\ Quant.\ Grav.\  {\bf 19} (2002) L87
  doi:10.1088/0264-9381/19/10/101
  [hep-th/0201081].
    
  
   

\bibitem{Kovacs:2015xha}
  S.~Kovacs, Y.~Sato and H.~Shimada,
  ``On membrane interactions and a three-dimensional analog of Riemann surfaces,''
  JHEP {\bf 1602} (2016) 050
  doi:10.1007/JHEP02(2016)050
  [arXiv:1508.03367 [hep-th]].

\bibitem{Kovacs:2013una}
  S.~Kovacs, Y.~Sato and H.~Shimada,
  ``Membranes from monopole operators in ABJM theory: Large angular momentum and M-theoretic $AdS_4/CFT_3$,''
  PTEP {\bf 2014} (2014) no.9,  093B01
  doi:10.1093/ptep/ptu102
  [arXiv:1310.0016 [hep-th]].


\bibitem{Yee:2003ge}
  J.~T.~Yee and P.~Yi,
  ``Instantons of M(atrix) theory in PP wave background,''
  JHEP {\bf 0302} (2003) 040
  doi:10.1088/1126-6708/2003/02/040
  [hep-th/0301120].   
  

\bibitem{RBNahm}
W.~Nahm, ``A Simple Formalism for the BPS Monopole'',
{\it Phys. Lett.} {\bf B90} (1980) 413; \\
W.~Nahm, ``All Selfdual Multi - Monopoles For Arbitrary Gauge Groups'',
CERN-TH-3172, C81-08-31.1-1.  

 

\bibitem{Ward:1989nz}
R.~S.~Ward,
``Linearization of the SU(infinity) Nahm Equations'',
{\it Phys. Lett.} {\bf B234} (1990) 81.

\bibitem{Hoppe:1994np}
J.~Hoppe,
``Surface motions and fluid dynamics'',
{\it Phys. Lett.} {\bf B335} (1994) 41
[hep-th/9405001].
  
  
 \bibitem{Sommerfeld:1896} 
 A.~Sommerfeld, ``\"Uber verzweigte Potentiate im Raum,'' 
 Proc.\ London\ Math.\ Soc.\ {\bf 28} (1896) 395.  
 
  \bibitem{Sommerfeld:1899} 
 A.~Sommerfeld,  
 Proc.\ London\ Math.\ Soc.\ {\bf 30} (1899) 161


\bibitem{SheikhJabbari:2004ik}
  M.~M.~Sheikh-Jabbari,
  ``Tiny graviton matrix theory: DLCQ of IIB plane-wave string theory, a conjecture,''
  JHEP {\bf 0409} (2004) 017
  doi:10.1088/1126-6708/2004/09/017
  [hep-th/0406214].



\bibitem{SheikhJabbari:2005mf}
  M.~M.~Sheikh-Jabbari and M.~Torabian,
  ``Classification of all 1/2 BPS solutions of the tiny graviton matrix theory,''
  JHEP {\bf 0504} (2005) 001
  doi:10.1088/1126-6708/2005/04/001
  [hep-th/0501001]. 


\bibitem{Basu:2004ed}
  A.~Basu and J.~A.~Harvey,
  ``The M2-M5 brane system and a generalized Nahm's equation,''
  Nucl.\ Phys.\ B {\bf 713} (2005) 136
  doi:10.1016/j.nuclphysb.2005.02.007
  [hep-th/0412310].  


\bibitem{Howe:1997ue}
P.~S.~Howe, N.~D.~Lambert and P.~C.~West,
``The Selfdual string soliton,'' Nucl.\ Phys.\ B {\bf 515} (1998) 203
  doi:10.1016/S0550-3213(97)00750-5
  [hep-th/9709014].


\bibitem{Niarchos:2012pn} 
  V.~Niarchos and K.~Siampos,
  ``M2-M5 blackfold funnels,''
  JHEP {\bf 1206}, 175 (2012)
  doi:10.1007/JHEP06(2012)175
  [arXiv:1205.1535 [hep-th]].

\bibitem{Ho:2008nn} 
  P.~M.~Ho and Y.~Matsuo,
  ``M5 from M2,''
  JHEP {\bf 0806}, 105 (2008)
  doi:10.1088/1126-6708/2008/06/105
  [arXiv:0804.3629 [hep-th]];
  P.~M.~Ho, Y.~Imamura, Y.~Matsuo and S.~Shiba,
  ``M5-brane in three-form flux and multiple M2-branes,''
  JHEP {\bf 0808}, 014 (2008)
  doi:10.1088/1126-6708/2008/08/014
  [arXiv:0805.2898 [hep-th]].
  
\bibitem{Lambert:2010wm} 
  N.~Lambert and C.~Papageorgakis,
  ``Nonabelian (2,0) Tensor Multiplets and 3-algebras,''
  JHEP {\bf 1008}, 083 (2010)
  doi:10.1007/JHEP08(2010)083
  [arXiv:1007.2982 [hep-th]];
  N.~Lambert and D.~Sacco,
  ``M2-branes and the (2, 0) superalgebra,''
  JHEP {\bf 1609}, 107 (2016)
  doi:10.1007/JHEP09(2016)107
  [arXiv:1608.04748 [hep-th]].

\bibitem{Chu:2009iv}   
  C.~S.~Chu and S.~L.~Ko,
  ``Non-abelian Action for Multiple Five-Branes with Self-Dual Tensors,''
  JHEP {\bf 1205}, 028 (2012)
  doi:10.1007/JHEP05(2012)028
  [arXiv:1203.4224 [hep-th]];
  C.~S.~Chu and D.~J.~Smith,
  ``Towards the Quantum Geometry of the M5-brane in a Constant C-Field from Multiple Membranes,''
  JHEP {\bf 0904}, 097 (2009)
  doi:10.1088/1126-6708/2009/04/097
  [arXiv:0901.1847 [hep-th]].
  
\bibitem{Saemann:2017zpd} 
  C.~Saemann and L.~Schmidt,
  ``Towards an M5-Brane Model I: A 6d Superconformal Field Theory,''
  arXiv:1712.06623 [hep-th].
  

  

\bibitem{Takayanagi:2002nv}
  H.~Takayanagi and T.~Takayanagi,
  ``Notes on giant gravitons on PP waves,''
  JHEP {\bf 0212} (2002) 018
  doi:10.1088/1126-6708/2002/12/018
  [hep-th/0209160].







\bibitem{Sadri:2003mx}
  D.~Sadri and M.~M.~Sheikh-Jabbari,
  ``Giant hedgehogs: Spikes on giant gravitons,''
  Nucl.\ Phys.\ B {\bf 687} (2004) 161
  doi:10.1016/j.nuclphysb.2004.03.013
  [hep-th/0312155].



\bibitem{Davis:1971}
L.~C.~Davis and J.~R.~Reitz, 
``Solution to Potential Problems near a Conducting Semi-Infinite Sheet or Conducting Disk,'' 
Am.\ J.\ Phys.\ {\bf 39}, (1971) 1225.    

\bibitem{Hobson:1900} 
E.~W.~Hobson, ``On Green's Function for a Circular Disc, with applications to Electro-static Problems,'' 
Trans.\ Camb.\ Phil.\ Soc.\ {\bf 18} (1900) 277. 


\bibitem{deMelloKoch:2016whh} 
  R.~de Mello Koch, D.~Gossman, L.~Nkumane and L.~Tribelhorn,
  ``Eigenvalue Dynamics for Multimatrix Models,''
  Phys.\ Rev.\ D {\bf 96}, no. 2, 026011 (2017)
  doi:10.1103/PhysRevD.96.026011
  [arXiv:1608.00399 [hep-th]];
  R.~de Mello Koch and L.~Nkumane,
  ``From Gauss Graphs to Giants,''
  JHEP {\bf 1802}, 005 (2018)
  doi:10.1007/JHEP02(2018)005
  [arXiv:1710.09063 [hep-th]].


\bibitem{Bachas:2000dx}
  C.~Bachas, J.~Hoppe and B.~Pioline,
  ``Nahm equations, N=1* domain walls, and D strings in AdS(5) x S(5),''
  JHEP {\bf 0107} (2001) 041
  doi:10.1088/1126-6708/2001/07/041
  [hep-th/0007067].
 



\bibitem{Constable:1999ac}
  N.~R.~Constable, R.~C.~Myers and O.~Tafjord,
  ``The Noncommutative bion core,''
  Phys.\ Rev.\ D {\bf 61} (2000) 106009
  doi:10.1103/PhysRevD.61.106009
  [hep-th/9911136].
  
  
\bibitem{Sakai:2013gga} 
  K.~Sakai and S.~Terashima,
  ``Integrability of BPS equations in ABJM theory,''
  JHEP {\bf 1311}, 002 (2013)
  doi:10.1007/JHEP11(2013)002
  [arXiv:1308.3583 [hep-th]].
  
 \bibitem{Guralnik:2000pb}
  Z.~Guralnik and S.~Ramgoolam,
  ``On the Polarization of unstable D0-branes into noncommutative odd spheres,''
  JHEP {\bf 0102} (2001) 032
  doi:10.1088/1126-6708/2001/02/032
  [hep-th/0101001].
  
  
  
  
  
\bibitem{Ramgoolam:2001zx}
  S.~Ramgoolam,
  ``On spherical harmonics for fuzzy spheres in diverse dimensions,''
  Nucl.\ Phys.\ B {\bf 610} (2001) 461
  doi:10.1016/S0550-3213(01)00315-7
  [hep-th/0105006].
    
 

  
  
  
  
  
  

    
  
  
  
  
  
  
  
  
  
  
  
  
  
  
  
  
  
  
  
  
  
  
  
  
  
  
  
    
\end{thebibliography}
\end{document}